\documentclass[aps,prd,showpacs,nofootinbib,floatfix,eqsecnum]{revtex4}  
\usepackage{amsmath,latexsym,graphicx,multirow,longtable}
\usepackage{verbatim}

\newcommand{\sech}{\mathrm{sech}}
\newcommand{\pder}[2]{\frac{\partial #1}{\partial #2}}

\newcommand{\Hcal}{\mathcal{H}}

\begin{document}
\title{Spacetime and orbits of bumpy black holes}
\author{Sarah J.\ Vigeland}
\author{Scott A.\ Hughes}
  
\affiliation{Department of Physics and MIT Kavli Institute,
Massachusetts Institute of Technology, 77 Massachusetts Avenue,
Cambridge, MA 02139}

\date{\today}  

\begin{abstract}
Our universe contains a great number of extremely compact and massive
objects which are generally accepted to be black holes.  Precise
observations of orbital motion near candidate black holes have the
potential to determine if they have the spacetime structure that
general relativity demands.  As a means of formulating measurements to
test the black hole nature of these objects, Collins and Hughes
introduced ``bumpy black holes'': objects that are almost, but not
quite, general relativity's black holes.  The spacetimes of these
objects have multipoles that deviate slightly from the black hole
solution, reducing to black holes when the deviation is zero.  In this
paper, we extend this work in two ways.  First, we show how to
introduce bumps which are smoother and lead to better behaved orbits
than those in the original presentation.  Second, we show how to make
bumpy Kerr black holes --- objects which reduce to the Kerr solution
when the deviation goes to zero.  This greatly extends the
astrophysical applicability of bumpy black holes.  Using
Hamilton-Jacobi techniques, we show how a spacetime's bumps are
imprinted on orbital frequencies, and thus can be determined by
measurements which coherently track the orbital phase of a small
orbiting body.  We find that in the weak field, orbits of bumpy black
holes are modified exactly as expected from a Newtonian analysis of a
body with a prescribed multipolar structure, reproducing well-known
results from the celestial mechanics literature.  The impact of bumps
on strong-field orbits is many times greater than would be predicted
from a Newtonian analysis, suggesting that this framework will allow
observations to set robust limits on the extent to which a spacetime's
multipoles deviate from the black hole expectation.
\end{abstract}
\maketitle

\section{Introduction}  
\label{sec:intro}

\subsection{Motivation: Precision tests of the black hole hypothesis}

Though observations of gravitational systems agree well with the
predictions of general relativity (GR), the most detailed and
quantitative tests have so far been done in the weak field.  (``Weak
field'' means that the dimensionless Newtonian potential $\phi \equiv
GM/rc^2 \ll 1$, where $M$ is a characteristic mass scale and $r$ a
characteristic distance.)  This is largely because many of the most
precise tests are done in our solar system (e.g., {\cite{bertotti}}).
Even the celebrated tests which use binary neutron stars (e.g.,
{\cite{pulsars}}) are essentially weak-field: for those systems, $M
\sim\,{\rm several}\,M_\odot$, $r = {\rm orbital\ separation}
\sim\,{\rm several}\,R_\odot$, so $\phi \sim\,{\rm a\ few}\times
GM_\odot/R_\odot c^2 \sim \, {\rm a\ few}\times 10^{-6}$.

This situation is on the verge of changing.  Observational technology
is taking us to a regime where we either are or soon will be probing
motion in strong gravity, with $\phi \gtrsim 0.1$.  Examples of
measurements being made now include radio studies of accretion flows
near the putative black hole in our galactic center {\cite{shep}} and
x-ray studies that allow precise measurements of accretion disk
geometries {\cite{psaltis_livrev,nms08}}.  Future measurements include
the possible discovery of a black hole-pulsar system, perhaps with the
Square Kilometer Array {\cite{smits}}, and gravitational-wave (GW)
observations of small bodies spiraling into massive black holes due to
the backreaction of GW emission {\cite{emri}}.

For weak-field studies, a well-developed paradigm for testing gravity
has been developed.  The {\it parameterized post-Newtonian} (PPN)
expansion {\cite{mtw, will93}} quantifies various measurable aspects
of relativistic gravity.  For example, the PPN parameter $\gamma$
(whose value is 1 in GR) quantifies the amount of spatial curvature
produced by a unit of rest mass.  Other PPN parameters quantify a
theory's nonlinearity, the degree to which it incorporates preferred
frames, and the possible violation of conservation laws.  See Ref.\
{\cite{will93}}, Chap.\ 4 for a detailed discussion.  Unfortunately, no
similar framework exists for strong-field studies.  If we hope to use
observations as tools for testing the nature of strong gravity objects
and strong-field gravity, we need to rectify this.

Black holes are of particular interest for studying strong-field
gravity.  Aside from having the strongest accessible gravitational
fields of any object in the universe\footnote{More accurately, they
have the largest potential $\phi \sim GM/Rc^2$.  One may also
categorize weak or strong gravity using spacetime curvature or tides;
arguably this is a more fundamental measure for assessing whether GR
is likely to be accurate or not.  From this perspective, black holes
are actually {\it not} such ``strong gravity'' objects; indeed, the
tidal field just outside a $10^8\,M_\odot$ black hole's event horizon
is not much different from the tidal field at the surface of the
Earth.  See Ref.\ {\cite{psaltis_livrev}} for further discussion of
this point.}, within GR they have an amazingly simple spacetime
structure: the ``no-hair'' theorems
{\cite{nohair1,nohair2,nohair3,nohair4,nohair5}} guarantee that the
exterior spacetime of any black hole is completely described by only
two numbers, its mass $M$ and spin parameter $a$.  {\it Any} deviation
from that simplicity points to a failure either in our understanding
of gravity or in the nature of ultracompact objects.

Recent work by Brink {\cite{brink1}} has reviewed in detail the
challenges involved in testing the spacetime structure of massive
compact bodies when one relaxes the assumption that the spacetime is
Kerr.  As in Brink's discussion, we will focus on spacetimes that are
stationary, axisymmetric, and vacuum in some exterior region.  An
especially useful parameterization is provided by the work of Geroch
{\cite{geroch70}} and Hansen {\cite{hansen74}}.  They demonstrate that
such a spacetime is completely specified by a set of mass multipole
moments $M_l$ and current multipole moments $S_l$.  For a fluid body,
$M_l$ describes the angular distribution of the body's mass, and $S_l$
describes the angular distribution of mass flow.  In Newtonian theory,
$M_l$ labels a piece of the gravitational potential that varies as
$Y_{l0}(\theta,\phi)/r^{l+1}$.  There is no Newtonian analog to $S_l$.
In the weak-field limit of GR, they label a magnetic-like contribution
to gravity and are reminiscent of magnetic moments.

In general, the moments $\{M_l, S_l\}$ can be arbitrary.  For black
holes, they take special values: a Kerr spacetime has
{\cite{hansen74}}
\begin{equation}
M_l + i S_l = M(ia)^l\;,
\label{eq:Kerr_moments}
\end{equation}
(in units with $G = 1 = c$; we neglect the
astrophysically uninteresting possibility of the hole having
macroscopic charge).  In other words, only two moments out of the set
$\{M_l, S_l\}$ are independent: $M_0 = M$, and $S_1 = aM$.  Once those
two moments have been specified, all other moments are fixed by Eq.\
(\ref{eq:Kerr_moments}) {\it if} the spacetime is Kerr.

Testing the hypothesis that an object is a Kerr black hole can thus be
framed as a null experiment.  First, measure the putative black hole
spacetime's multipoles.  Using the moments $M_0$ and $S_1$, determine the
parameters $M$ and $a$.  {\it If the spacetime is Kerr, all moments
for $l \ge 2$ must be given by Eq.\ (\ref{eq:Kerr_moments}).}  The
null hypothesis is that any deviation from those Kerr moments is zero.
Failure of the null hypothesis means the black hole candidate is not a
Kerr black hole, and may indicate a failure of strong-field GR.

\subsection{Bumpy black holes: Past work}

Performing this null experiment requires strong-field spacetimes with
multipoles differing from those of black holes.  Given the tremendous
success of general relativity and the black hole hypothesis at
explaining a wide span of data, a reasonable starting point is to
consider spacetimes whose multipoles deviate only slightly.  Imagine
starting with a spacetime whose Geroch-Hansen moments satisfy
\begin{equation}
M_l + i S_l = M(ia)^l + \delta M_l + i\delta S_l\;.
\label{eq:Bumpy_Kerr_moments}
\end{equation}
The null hypothesis --- black hole candidates are described by GR's
Kerr metric --- means $\delta M_l = 0$, $\delta S_l = 0$ for all $l$.

To this end, Collins and Hughes {\cite{ch04}} (hereafter CH04)
introduced the {\it bumpy black hole}: a spacetime that deviates in a
small, controllable manner from the exact black holes of GR.  By
construction, the bumpy black hole includes ``normal'' black holes as
a limit.  This is central to testing the black hole hypothesis by a null
experiment.  Spacetimes of other proposed massive, compact objects
(for example, boson stars {\cite{csw86, ryan97}}) typically do not
include black holes as a limiting case.  This limits their utility if
black hole candidates are, in fact, GR's black holes.  Measurements of
black holes using observables formulated in a bumpy black hole
spacetime should simply measure the spacetime's ``bumpiness'' (defined
more precisely below) to be zero.

Though a useful starting point, the bumpy black holes developed in
CH04 had two major shortcomings.  First, the changes to the spacetime
that were introduced to modify its multipole moments were not smooth.
The worked example presented in CH04 is interpreted as a Schwarzschild
black hole perturbed by an infinitesimally thin ring of positive mass
around its equator and by a pair of negative mass infinitesimal
points near its poles.  Though this changes the spacetime's quadrupole
moment (the desired outcome of this construction), it gives the
spacetime a pathological strong-field structure.  This is reflected in
the fact that non-equatorial strong-field orbits are ill-behaved in
the CH04 construction {\cite{dimitrios_private}}.

Second, CH04 only examined bumpy Schwarzschild black holes.
Although their approach differed in many details from that used in
CH04, Glampedakis and Babak {\cite{gb06}} (hereafter GB06) rectified
this deficiency with their introduction of a ``quasi-Kerr'' spacetime.
Their construction uses the exterior Hartle-Thorne metric
{\cite{h67,ht68}} describing the exterior of any slowly rotating,
axisymmetric, stationary body.  It includes the Kerr metric to ${\cal
O}(a^2)$ as a special case.  Identifying the influence of the
quadrupole moment in the Hartle-Thorne form of the Kerr metric, they
then introduce a modification to the full Kerr spacetime that changes
the black hole's quadrupole moment from its canonical value.

\subsection{This analysis}

The goal of this work is to rectify the deficiencies of CH04 and to
extend the bumpy black hole concept to spinning black holes.  Dealing
with the non-smooth nature of the bumps presented in CH04 is, as we
show in Secs.\ {\ref{sec:spacetimes}}, {\ref{sec:results_schw}}, and
{\ref{sec:results_kerr}}, quite straightforward.  It simply requires
introducing perturbations to the black hole background that are smooth
rather than discontinuous.  In essence, rather than having bumps that
correspond to infinitesimal points and rings, the bumps we use here
are smeared into pure multipoles.

Extending CH04 to spinning black holes is more of a challenge.  Given
that GB06 already introduced a bumpy-black-hole-like spacetime that
encompasses spacetimes with angular momentum, one might wonder why
another construction is needed.  A key motivation is that we would
like to be able to make an {\it arbitrary} modification to a black
hole's moments.  Although showing that a black hole candidate has a
non-Kerr value for the quadrupole moment would be sufficient to
falsify its black hole nature (at least within the framework of GR),
one can imagine scenarios in which the first $L$ moments of a black
hole candidate agree with the Kerr value, but things differ for $l >
L$.  For example, Yunes et al.\ have shown that in Chern-Simons
modifications to GR, slowly rotating black hole solutions have the
multipolar structure of Kerr for $l < 4$, but differ for $l \ge 4$
{\cite{csbh}}.  There are many ways in which black hole candidates
might differ from the black holes of general relativity; we need to
develop a toolkit sufficiently robust that it can encompass these many
potential points of departure.

Our technique for making bumpy Kerr holes is based on the Newman-Janis
algorithm {\cite{nj}}.  This algorithm transforms a Schwarzschild
spacetime into Kerr by ``rotating'' the spacetime in a complex
configuration space\footnote{This complex ``rotation'' is made most
clear by framing the discussion using the Ernst potential
{\cite{ernst}}.  Transforming from Schwarzschild to Kerr corresponds
to adding an imaginary part to a particular potential.  Further
discussion is given in a companion paper {\cite{vigeland}}.}.  In this
paper, we construct bumpy Kerr black holes by applying the
Newman-Janis algorithm to bumpy Schwarzschild black holes.  The
outcome of this procedure is a spacetime whose mass moments are
deformed relative to Kerr.  When the bumpiness is set equal to zero,
we recover the Kerr metric.  A companion paper {\cite{vigeland}}
examines the multipolar structure of bumpy Kerr black holes in more
detail, demonstrating that one can construct a spacetime in which the
mass moments deviate arbitrarily from Kerr (provided that they are
small).

Once one has constructed a bumpy black hole spacetime, one then needs
to show how its bumps are encoded in observables.  The most detailed
quantitative tests will come from orbits near black hole candidates.
As such, it is critical to know how orbital frequencies change as a
function of a spacetime's bumpiness.  More generally, we are faced
with the problem of understanding motion in general stationary
axisymmetric vacuum spacetimes.  Brink {\cite{brink2}} has recently
published a very detailed analysis of this problem, with a focus on
understanding whether and for which situations the spacetimes admit
integrable motion.  She has found evidence that geodesic motion in
such spacetimes may, in many cases, be integrable.  If so, the problem
of mapping general spacetimes (not just ``nearly black hole''
spacetimes) may be tractable.  Gair, Li, and Mandel {\cite{glm08}}
have similarly examined orbital characteristics in the Manko-Novikov
spacetime {\cite{mn92}}, which has a particular tunable non-Kerr
structure.  They show how orbits change in such spacetimes, and how
its bumpiness colors observable characteristics.

For this analysis, we confine ourselves to the simpler problem of
motion in bumpy black hole spacetimes, addressing this challenge in
Secs.\ {\ref{sec:results_schw}} and {\ref{sec:results_kerr}} using
canonical perturbation theory.  As is now well known (and was shown
rather spectacularly by Schmidt {\cite{schmidt}}), Hamilton-Jacobi
methods let us write down closed-form expressions for the three
orbital frequencies ($\Omega^r$, $\Omega^\theta$, $\Omega^\phi$) which
completely characterize the behavior of bound Kerr black hole orbits.
Since bumpy black hole spacetimes differ only perturbatively from
black hole spacetimes, canonical perturbation theory lets us
characterize how a spacetime's bumps shift those frequencies, and thus
are encoded in observables.  Similar techniques were used in GB06 to
see how frequencies are shifted in a quasi-Kerr metric, and were also
used by Hinderer and Flanagan {\cite{hf08}} in a two-timescale
analysis of inspiral into Kerr black holes.

\subsection{Organization and overview}

We begin in Sec.\ {\ref{sec:spacetimes}} with an overview of the
spacetimes that we study.  We start with the axially symmetric and
stationary Weyl line element.  We first review the Einstein field
equations in this representation, introduce the Schwarzschild limit,
and then describe first order perturbations.  The spacetime's
bumpiness is set by choosing a function $\psi_1$ which controls how
the spacetime deviates from the black hole limit.  We initially leave
$\psi_1$ arbitrary, except for the requirement that it be small enough
that terms of order $(\psi_1)^2$ can be neglected.  Later in the
paper, we will take $\psi_1$ to be a pure multipole in the Weyl
representation.  Following our discussion of bumpy Schwarzschild
spacetimes, we show how to use the Newman-Janis algorithm to build
bumpy Kerr black holes.

We study geodesic motion in these spacetimes in Sec.\
{\ref{sec:motion}}.  We begin by reviewing the most important
properties of normal black hole orbits, reviewing Kerr geodesics in
Sec.\ {\ref{sec:kerr_geodesics}}, and then describing how to compute
orbital frequencies using Hamilton-Jacobi methods in Sec.\
{\ref{sec:normal_bh_freqs}}.  The discussion of frequencies is largely
a synopsis of Schmidt's pioneering study, Ref.\ {\cite{schmidt}}.  We
then show how these techniques can be adapted, using canonical
perturbation theory, to bumpy black holes (Sec.\
{\ref{sec:bumpy_bh_freqs}}).  Canonical perturbation theory requires
averaging a bump's shift to an orbit's Hamiltonian.  This averaging
was developed in Ref.\ {\cite{dh04}}, and is summarized in Appendix
{\ref{app:averaging}}.

We present our results for bumpy Schwarzschild and bumpy Kerr orbits
in Secs.\ {\ref{sec:results_schw}} and {\ref{sec:results_kerr}},
respectively.  We take $\psi_1$ to be a pure multipole in the Weyl
sector, construct the spacetime, and numerically compute the shifts in
$\Omega^r$, $\Omega^\theta$, and $\Omega^\phi$.  Detailed results are
given for $l = 2$, $l = 3$, and $l = 4$.  There is no reason in
principle to stop there, though the results quickly become repetitive.
Section {\ref{sec:results_schw}} gives our results for bumpy
Schwarzschild black holes, and Sec.\ {\ref{sec:results_kerr}} gives
results for bumpy Kerr.

A few results are worth highlighting.  First, we find that the exact
numerical results for frequency shifts correctly reproduce the
Newtonian limit as our orbits are taken into the weak field.  (We
develop this limit in some detail in Appendix {\ref{app:newton}} to
facilitate the comparison.)  The frequency shifts are substantially
enhanced in the strong field.  The shift to the radial frequency is
particularly interesting: it tends to oscillate, shifting between an
enhancement and a decrement as orbits move into the strong field.
This behavior appears to be a robust signature of non-Kerr multipole
structure in black hole strong fields.

Interestingly, it turns out that black hole spin does not have a very
strong impact on the bumpiness-induced shifts to orbital frequencies.
Spin's main effect is to change the location of the last stable orbit.
For large spin, orbits reach deeper into the strong field, amplifying
the bumps' impact on orbital frequencies.  Aside from the change to
the last stable orbits, the impact of a particular multipolar bump
looks largely the same across all spin values.  Some examples of the
frequency shifts we find are shown in Figs.\ {\ref{fig:dOm_l2}} ($l =
2$, Schwarzschild), {\ref{fig:dOm_l4}} ($l = 4$, Schwarzschild) and
{\ref{fig:dOm_l2_Kerr}} ($l = 2$, Kerr).  (We have no figures for $l =
3$ since the secular shifts to orbital frequencies are zero in that
case, as they are for all odd values of $l$.  We also do not show
results for $l = 4$ bumps of Kerr black holes since, as discussed
above, they are very similar to the $l = 4$ Schwarzschild results.)

Finally, we summarize our analysis and suggest some directions for
future work in Sec.\ {\ref{sec:summary}}.  Among the points we note
are that, in this analysis, we only consider changes to the mass
moments of the black hole spacetime.  Adding ``current-type'' bumps to
a spacetime is discussed in a companion paper {\cite{vigeland}}.  We
also do not discuss in this analysis the issue of measurability.
Turning these foundations for mapping the multipole moment structure
of black holes into a practical measurement program (for instance, via
gravitational-wave measurements, timing of a black hole-pulsar binary,
or precision mapping in radio or x-rays of accretion flows) will take
a substantial effort.

Throughout this paper, we work in geometrized units with $G = c = 1$; a
useful conversion factor in these units is $1\,M_\odot = 4.92\times
10^{-6}$ seconds.  When we discuss bumpy black holes, we will always
use a ``hat'' accent to denote quantities which are calculated in the
pure black hole background spacetimes.  For example, an orbital
frequency is written $\Omega = \hat\Omega + \delta\Omega$;
$\hat\Omega$ is the frequency of an orbit in the black hole
background, and $\delta\Omega$ denotes the shift due to the black
hole's bumps.

\section{Black hole and bumpy black hole spacetimes}
\label{sec:spacetimes}

We begin with general considerations on the spacetimes we consider.
Since we focus on stationary, axisymmetric spacetimes, the Weyl metric
{\cite{weyl}} is a good starting point:
\begin{equation}
ds^2 = -e^{2\psi}dt^2 + e^{2\gamma-2\psi}(d\rho^2+dz^2)
+ e^{-2\psi}\rho^2d\phi^2 \;.
\label{eq:weyl}
\end{equation}
The nontrivial vacuum Einstein equations for this metric are given by
\begin{eqnarray}
0 &=& 
\frac{\partial^2\psi}{\partial\rho^2} +
\frac{1}{\rho}\frac{\partial\psi}{\partial\rho} +
\frac{\partial^2\psi}{\partial z^2}\;,
\label{eq:weyl_einstein1}
\\
\frac{\partial\gamma}{\partial\rho} &=&
\rho\left[\left(\frac{\partial\psi}{\partial\rho}\right)^2 -
\left(\frac{\partial\psi}{\partial z}\right)^2\right] \;,
\label{eq:weyl_einstein2}
\\
\frac{\partial\gamma}{\partial z} &=& 2\rho
\frac{\partial\psi}{\partial\rho}\frac{\partial\psi}{\partial z} \;.
\label{eq:weyl_einstein3}
\end{eqnarray}
Equations (\ref{eq:weyl_einstein1}) -- (\ref{eq:weyl_einstein3}) will
be our main tools for building bumpy black hole spacetimes.  We will
put $\psi = \psi_0 + \psi_1$, $\gamma = \gamma_0 + \gamma_1$, with
$\psi_1/\psi_0 \ll 1$, and $\gamma_1/\gamma_0 \ll 1$.  Before
specializing to black hole backgrounds, note that Eq.\
(\ref{eq:weyl_einstein1}) is simply Laplace's equation.  The functions
$\psi_1$ can thus very conveniently be taken to be harmonic functions.
This is key to smoothing out the spacetime's bumps and curing one of
the deficiencies of CH04.

\subsection{Schwarzschild and bumpy Schwarzschild}

We begin by building a bumpy Schwarzschild black hole.  The
Schwarzschild metric is recovered from Eq.\ (\ref{eq:weyl}) when
$\psi_1 = \gamma_1 = 0$ and we set
\begin{eqnarray}
\psi_0 &=& \ln \tanh(u/2)\;,
\label{eq:psi_schw}
\\
\gamma_0 &=& -\frac{1}{2}\ln\left(1 + \frac{\sin^2v}{\sinh^2u}\right)\;.
\label{eq:gamma_schw}
\end{eqnarray}
The prolate spheroidal coordinates $(u,v)$ are a remapping of the
coordinates $(\rho,z)$ used in Eq.\ (\ref{eq:weyl}):
\begin{eqnarray}
\rho &=& M \sinh u \sin v\;,
\label{eq:rho_of_uv}
\\
z &=& M \cosh u \cos v \;.
\label{eq:z_of_uv}
\end{eqnarray}
(Below we will remap these to the familiar Schwarzschild form.)
Expanding the Einstein equations Eqs.\ (\ref{eq:weyl_einstein1}) --
(\ref{eq:weyl_einstein3}) about these Schwarzschild values to leading
order in $\psi_1$ and $\gamma_1$, the perturbations must satisfy
\begin{eqnarray}
\nabla^2\psi_1 &=& 0\;,
\label{eq:laplace_psi1}
\\
\frac{\partial\gamma_1}{\partial u} &=& \frac{2[\tan
v(\partial\psi_1/\partial u) + \tanh u(\partial\psi_1/\partial
v)]}{\sinh u(\coth u \tan v + \tanh u \cot v)} \;,
\label{gamma_constraint1}
\\
\frac{\partial\gamma_1}{\partial v} &=& \frac{2[\tan
v(\partial\psi_1/\partial v) - \tanh u(\partial\psi_1/\partial
u)]}{\sinh u(\coth u \tan v + \tanh u \cot v)}\;.
\label{gamma_constraint}
\end{eqnarray}
As discussed in CH04, Eqs.\ (\ref{gamma_constraint1}) and
(\ref{gamma_constraint}) actually overdetermine the solution; we will
use Eq.\ (\ref{gamma_constraint}) to calculate $\gamma_1$.  [Note also
that the $\tan v$ in the numerator of Eq.\ (\ref{gamma_constraint1})
is incorrectly written $\cot v$ in CH04.]

To connect this to the variables we will use later in the paper (and
to put it in a more familiar form), we make a final change of
coordinates, putting
\begin{eqnarray}
r &=& 2M \cosh^2(u/2) \;,
\label{eq:r_of_u}
\\
\theta &=& v \;,
\label{eq:theta_of_v}
\end{eqnarray}
so that
\begin{eqnarray}
\rho &=& r \sin\theta \sqrt{1 - \frac{2M}{r}} \;, \\
z &=& (r - M)\cos\theta \;.
\end{eqnarray}
The spacetime then becomes
\begin{eqnarray}
ds^2 &=& -e^{2\psi_1}\left(1 - \frac{2M}{r}\right)dt^2
+ e^{2\gamma_1 - 2\psi_1}\left(1 - \frac{2M}{r}\right)^{-1}dr^2
+ r^2e^{2\gamma_1 - 2\psi_1}d\theta^2 + r^2\sin^2\theta
e^{-2\psi_1}d\phi^2
\nonumber\\
&\equiv& \left(\hat g_{\alpha\beta} + b_{\alpha\beta}\right)
dx^\alpha dx^\beta\;.
\label{perturbSchw}
\end{eqnarray}
Although we have left the potentials $\psi_1$ and $\gamma_1$ in
exponential form, these quantities must be expanded to first order,
since we solve for them using linearized Einstein equations.  We use
the exponential form only for notational convenience.  On the second
line, $\hat g_{\alpha\beta}$ is the Schwarzschild metric, and
\begin{eqnarray}
b_{tt} &=& -2\psi_1\left(1 - \frac{2M}{r}\right)\;,
\label{eq:btt_schw}\\
b_{rr} &=& \left(2\gamma_1 - 2\psi_1\right)\left(1 -
\frac{2M}{r}\right)^{-1}\;,
\label{eq:brr_schw}\\
b_{\theta\theta} &=& \left(2\gamma_1 - 2\psi_1\right)r^2\;,
\label{eq:bthth_schw}\\
b_{\phi\phi} &=& -2\psi_1r^2\sin^2\theta\;.
\label{eq:bphph_schw}
\end{eqnarray}
All other components of $b_{\alpha\beta}$ are zero.  We clearly
recover the normal Schwarzschild black hole when $\psi_1 \to 0$,
$\gamma_1 \to 0$.

\subsection{Kerr and bumpy Kerr}
We now use the Newman-Janis algorithm {\cite{nj}} to transform bumpy
Schwarzschild into bumpy Kerr.  We begin with the bumpy Schwarzschild
metric written in prolate spheroidal coordinates:
\begin{equation}
ds^2 = -e^{2\psi_1}\tanh^2(u/2) dt^2 + e^{2\gamma_1-2\psi_1}
4M^2\cosh^4(u/2) (du^2 + dv^2) + e^{-2\psi_1}4 M^2\cosh^4(u/2)\sin^2v \,d\phi^2
\;.
\label{eq:bumpySchw_prolate}
\end{equation}
The first step in the Newman-Janis algorithm uses the fact that the
metric can be written in terms of a complex null tetrad with legs
$l^\mu$, $n^\nu$, $m^\nu$:
\begin{eqnarray}
g^{\mu\nu} &=& -l^\mu n^\nu - l^\nu n^\mu + m^\mu \bar{m}^\nu + m^\nu
\bar{m}^\mu\;;
\label{eq:metric_from_tetrad}
\end{eqnarray}
an overbar denotes complex conjugate.  The legs are given by
\begin{eqnarray}
l^\mu &=& e^{-\psi_1}\coth^2(u/2)\,\delta^\mu_t + \frac{1}{M}e^{\psi_1
- \gamma_1}{\text{csch}}\, u\, \delta^\mu_u \;,
\label{eq:l_leg}
\\
n^\mu &=& \frac{1}{2} e^{-\psi_1}\delta^\mu_t - \frac{1}{2M} e^{\psi_1
- \gamma_1}{\text{csch}}\,u\,\tanh^2(u/2)\,\delta^\mu_u \;,
\label{eq:n_leg}
\\
m^\mu &=& \frac{1}{2\sqrt{2}M}e^{\psi_1}\sech^2(u/2)
\left(e^{-\gamma_1}\,\delta^\mu_v + i\csc v \,\delta^\mu_\phi\right) \;.
\label{eq:m_leg}
\end{eqnarray}
We use the Kronecker delta $\delta^\mu_\nu$ to indicate components.
Next follows the key step of the Newman-Janis algorithm: We allow the
coordinate $u$ to be complex, and rewrite $l^\mu$, $n^\mu$, and
$m^\mu$ as
\begin{eqnarray}
l^\mu &=& e^{-\psi_1}\frac{2\;\delta^\mu_t}{\tanh^2(u/2) +
\tanh^2(\bar u/2)} + e^{\psi_1 -
\gamma_1}\frac{\delta^\mu_u}{M\sqrt{\cosh u\cosh\bar{u} - 1}} \;,
\\
n^\mu &=& \frac{1}{2} e^{-\psi_1}\,\delta^\mu_t - e^{\psi_1 - \gamma_1}
\frac{\left[\tanh^2(u/2) + \tanh^2(\bar{u}/2)\right]}
{4M\sqrt{\cosh u\cosh\bar{u} - 1}}\,\delta^\mu_u \;,
\\
m^\mu &=& \frac{1}{2\sqrt{2}M}e^{\psi_1}
\sech^2(\bar{u}/2)\left(e^{-\gamma_1}\,\delta^\mu_v + i\csc v \,
\delta^\mu_\phi\right) \;.
\end{eqnarray}
Notice that we recover the original tetrad when we force $u =
\bar{u}$.  Further discussion of this seemingly {\it ad hoc} procedure
(and an explanation of how it uniquely generates the Kerr spacetime)
is given in Ref.\ {\cite{ds00}}.

Next, change coordinates: Rewrite the tetrad using
$(U,r,\theta,\phi)$, given by
\begin{eqnarray}
U &=& t - 2M\cosh^2(u/2) - 2M\ln\left[\sinh^2(u/2)\right] -
ia\cos\theta \;,
\\
r &=& 2M \cosh^2(u/2) + ia\cos\theta \;,
\\
\theta &=& v \;.
\end{eqnarray}
The axial coordinate $\phi$ is the same in both coordinate systems.  At this
point, $a$ is just a parameter.  The result of this transformation is
\begin{eqnarray}
l^\mu &=& \left(e^{-\psi_1}-e^{\psi_1 - \gamma_1}\right) \left(1 -
\frac{2Mr}{\Sigma}\right)^{-1} \delta^\mu_U + e^{\psi_1 - \gamma_1}
\delta^\mu_r \;,
\\
n^\mu &=& \frac{1}{2}\left(e^{-\psi_1} + e^{\psi_1 -
\gamma_1}\right)\,\delta^\mu_U - \frac{1}{2}e^{\psi_1 - \gamma_1}
\left(1 - \frac{2Mr}{\Sigma}\right) \, \delta^\mu_r \;,
\\
m^\mu &=& \frac{e^{\psi_1 - \gamma_1}}{\sqrt{2}(r + ia\cos\theta)}
\left[ia\sin\theta\left(\delta^\mu_U - \delta^\mu_r\right) +
\delta^\mu_\theta + e^{\gamma_1}i\csc\theta\;\delta^\mu_\phi\right] \;,
\end{eqnarray}
where $\Sigma = r^2 + a^2\cos^2\theta$.  Making one further coordinate
transformation,
\begin{equation}
dU = dt - \frac{r^2 + a^2}{\Delta}dr \;,
\qquad
d\phi = d\phi' - \frac{a}{\Delta}dr \;,
\end{equation}
gives us a bumpy Kerr black hole metric in Boyer-Lindquist coordinates:
\begin{eqnarray}
ds^2 &=& -e^{2\psi_1}\left(1 - \frac{2Mr}{\Sigma}\right)dt^2
+ e^{2\psi_1 - \gamma_1}(1 -
e^{\gamma_1})\frac{4a^2Mr\sin^2\theta}{\Delta\Sigma}\;dt\;dr -
e^{2\psi_1 - \gamma_1}\frac{4aMr\sin^2\theta}{\Sigma}\;dt\;d\phi
\nonumber \\
&& + e^{2\gamma_1 - 2\psi_1}\left(1 - \frac{2Mr}{\Sigma}\right)^{-1}
\left[1 + e^{-2\gamma_1}(1 -
2e^{\gamma_1})\frac{a^2\sin^2\theta}{\Delta} - e^{4\psi_1 -
4\gamma_1}(1 -
e^{\gamma_1})^2\frac{4a^4M^2r^2\sin^4\theta}{\Delta^2\Sigma^2}\right]
dr^2
\nonumber \\
&& -2(1 - e^{\gamma_1})a\sin^2\theta \left[e^{-2\psi_1}\left(1 -
\frac{2Mr}{\Sigma}\right)^{-1} - e^{2\psi_1 -
2\gamma_1}\frac{4a^2M^2r^2\sin^2\theta}{\Delta\Sigma(\Sigma -
2Mr)}\right]\;dr\;d\phi
\nonumber\\
&& + e^{2\gamma_1 - 2\psi_1}\Sigma\;d\theta^2 + \Delta
\left[e^{-2\psi_1}\left(1 - \frac{2Mr}{\Sigma}\right)^{-1} -
e^{2\psi_1 - 2\gamma_1}\frac{4 a^2 M^2 r^2
\sin^2\theta}{\Delta\Sigma(\Sigma - 2Mr)}\right]
\;\sin^2\theta\;d\phi^2
\label{KerrBBH}
\end{eqnarray}
where $\Delta \equiv r^2-2Mr+a^2$, and we have dropped the prime on
$\phi$.  Notice that Eq.\ (\ref{KerrBBH}) reduces to the ``normal''
Kerr black hole metric when $\psi_1 \to 0$, $\gamma_1 \to 0$; the
parameter $a$ is seen to be the specific spin angular momentum, $|\vec
S|/M$.  A companion paper {\cite{vigeland}} examines the multipoles of
this spacetime for particular choices of $\psi_1$ and demonstrates
that it corresponds to Kerr with some moments set to the ``wrong''
values.  The Newman-Janis algorithm applied to the bumpy Schwarzschild
black hole produces a bumpy Kerr black hole.

Writing the bumpy Kerr metric in the form $g_{\alpha\beta} = {\hat
g}_{\alpha\beta} + b_{\alpha\beta}$, we read out of Eq.\
(\ref{KerrBBH})
\begin{eqnarray}
b_{tt} &=& -2\left(1-\frac{2Mr}{\Sigma}\right)\psi_1\;,
\label{eq:btt_Kerr}\\
b_{tr} &=& -\gamma_1\frac{2a^2Mr\sin^2\theta}{\Delta\Sigma}\;,
\label{eq:btr_Kerr}\\
b_{t\phi} &=&\left(\gamma_1-2\psi_1\right)
\frac{2aMr\sin^2\theta}{\Sigma}\;,
\label{eq:btph_Kerr}\\
b_{rr} &=& 2\left(\gamma_1-\psi_1\right)\frac{\Sigma}{\Delta}\;,
\label{eq:brr_Kerr}\\
b_{r\phi} &=& \gamma_1\left[\left(1-\frac{2Mr}{\Sigma}\right)^{-1}
-\frac{4a^2M^2r^2\sin^2\theta}{\Delta\Sigma(\Sigma-2Mr)}\right]
a\sin^2\theta\;,
\label{eq:brph_Kerr}\\
b_{\theta\theta} &=& 2\left(\gamma_1-\psi_1\right)\Sigma
\label{eq:bthth_Kerr}\\
b_{\phi\phi} &=& \left[ \left(\gamma_1-\psi_1\right)
\frac{8a^2M^2r^2\sin^2\theta}{\Delta\Sigma(\Sigma-2Mr)}
-2\psi_1\left(1-\frac{2Mr}{\Sigma}\right)^{-1} \right]
\Delta\sin^2\theta\;.
\label{eq:bphph_Kerr}
\end{eqnarray}
Other components are related by symmetry or zero.  By inspection, we
can see that $b_{\alpha\beta} \to 0$ as $\psi_1 \to 0$, $\gamma_1 \to
0$.

Before moving on, we summarize.  To build a bumpy black hole
spacetime, we first select a function $\psi_1$ which satisfies the
Laplace equation (\ref{eq:weyl_einstein1}).  We find the function
$\gamma_1$ which satisfies Eqs.\ (\ref{eq:weyl_einstein2}) and
(\ref{eq:weyl_einstein3}), and then apply the Newman-Janis algorithm
to ``rotate'' the spacetime to non-zero $a$.  The result is given by
Eqs.\ (\ref{KerrBBH}) -- (\ref{eq:bphph_Kerr}).

\section{Motion in black hole and bumpy black hole spacetimes}
\label{sec:motion}

We now discuss motion in these spacetimes.  Our focus will be
computing the frequencies associated with oscillations in the radial
coordinate $r$, the polar angle $\theta$, and rotations in $\phi$
about the symmetry axis.  These frequencies are typically the direct
observables of black hole orbits; it is from measuring these
frequencies (or the evolution of these frequencies if the orbit
evolves) that one can hope to constrain the properties of black hole
candidates.

\subsection{Geodesics of Kerr black holes}
\label{sec:kerr_geodesics}

As background to our discussion of black hole orbital frequencies, we
first briefly review the equations governing black hole geodesic
orbits, and some useful reparameterizations for practical
computations.  One goal of this discussion is to introduce certain
quantities which we will use in the remainder of this paper.

As first recognized by Carter {\cite{carter68}}, geodesic motion for a
test mass $m$ in a black hole spacetime is separable with respect to
Boyer-Lindquist coordinates $t$, $r$, $\theta$, and $\phi$.  The test
body's motion is then completely described by four first-order
ordinary differential equations:
\begin{eqnarray}
m^2\Sigma^2 \left(\frac{dr}{d\tau}\right)^2 &=& \left[(r^2 + a^2)E - a
L_z\right]^2 - \Delta\left[m^2 r^2 + (L_z - a E)^2 + Q\right]
\nonumber\\
&\equiv& R(r)\;,
\label{eq:r_of_tau}\\
m^2\Sigma^2 \left(\frac{d\theta}{d\tau}\right)^2 &=& Q - \cot^2\theta L_z^2
-a^2\cos^2\theta(m^2 -E^2)\;,
\nonumber\\
&\equiv& \Theta(\theta)\;,
\label{eq:theta_of_tau}\\
m\Sigma \left(\frac{d\phi}{d\tau}\right) &=&
\csc^2\theta\,L_z + aE\left(\frac{r^2 + a^2}{\Delta} - 1\right)
-\frac{a^2 L_z}{\Delta}
\nonumber\\
&\equiv& \Phi(r,\theta)\;,
\label{eq:phi_of_tau}\\
m\Sigma \left(\frac{dt}{d\tau}\right) &=&
E\left[\frac{(r^2 + a^2)^2}{\Delta} - a^2\sin^2\theta\right]
+ a L_z\left(1 - \frac{r^2 + a^2}{\Delta}\right)
\nonumber\\
&\equiv& T(r,\theta)\;.
\label{eq:t_of_tau}
\end{eqnarray}
As in the previous section, $\Delta = r^2 - 2Mr + a^2$ and $\Sigma =
r^2 + a^2\cos^2\theta$; $\tau$ is proper time along the test body's
worldline.  In developing these equations, one isolates four constants of
the motion.  One is simply the rest mass itself, $m^2 = -p^\mu p_\mu$;
this motivates the definition of the Hamiltonian for test body motion,
\begin{equation}
\Hcal \equiv \frac{1}{2} g^{\alpha\beta} p_\alpha p_\beta\;,
\label{eq:Hamiltonian_def}
\end{equation}
where the 4-momentum components $p_\mu = m g_{\mu\nu}(dx^\nu/d\tau)$.
The other constants are the energy $E$, axial angular momentum $L_z$,
and ``Carter constant'' $Q$, given by
\begin{eqnarray}
E &\equiv& -p_t\;,
\label{eq:Edef}\\
L_z &\equiv& p_\phi\;,
\label{eq:Lzdef}\\
Q &\equiv& p_\theta^2 + \cos^2\theta\left[a^2(m^2 - E^2) +
\csc^2\theta L_z^2\right]\;.
\label{eq:Qdef}
\end{eqnarray}
Given a choice of the constants $(E,L_z,Q)$ and a set of initial
conditions, Eqs.\ (\ref{eq:r_of_tau}) -- (\ref{eq:t_of_tau})
completely describe the geodesic motion of a test body near a Kerr
black hole.  The equations for $r$ and $\theta$ can present some
problems, however, since their motion includes turning points where
$dr/d\tau$ and $d\theta/d\tau$ pass through zero and switch sign.  To
account for this behavior, it is convenient to reparameterize these
motions using angles $\psi_r$ (for the radial motion) and $\chi$ (for
the polar motion) which smoothly vary from $0$ to $2\pi$ as the motion
oscillates between its extremes.

Consider first the radial motion.  We define
\begin{equation}
r = \frac{pM}{1 + e\cos\psi_r}\;.
\label{eq:r_of_psi}
\end{equation}
The constants $p$ and $e$ are the orbit's semi-latus rectum and
eccentricity, respectively.  Substituting into Eq.\
(\ref{eq:r_of_tau}), it is simple to develop an equation governing
$\psi_r$.  Periapsis and apoapsis are given by $r_p = pM/(1 + e)$ and
$r_a = pM/(1 - e)$, respectively.  This allows us to relate the
constants $p$ and $e$ to the constants $E$, $L_z$, and $Q$: They are
the outermost radii for which the radial ``potential'' goes to zero,
\begin{equation}
R(r_p) = R(r_a) = 0\;.
\label{eq:radial_turning_points}
\end{equation}

For the polar motion, we note that Eq.\ (\ref{eq:theta_of_tau}) can be
written
\begin{equation}
\Sigma^2 \left(\frac{d\theta}{d\tau}\right)^2 = \frac{z^2\left[a^2(m^2 -
E^2)\right] - z\left[Q + L_z^2 + a^2(m^2 - E^2)\right] + Q}{1 - z}\;,
\label{eq:theta_of_tau_2}
\end{equation}
where we have introduced $z \equiv \cos^2\theta$.  Denote by $z_{\pm}$
the two roots of the quadratic on the right-hand side of Eq.\
(\ref{eq:theta_of_tau_2}).  Turning points of the $\theta$ motion
correspond to $z = z_-$, the smaller of these roots.  (The root $z_+$
is greater than 1, and does not correspond to a turning point.)
Transforming back to the angle $\theta$, we find that the minimum
polar angle reached by the orbit is
\begin{equation}
\cos\theta_{\rm min} = \sqrt{z_-}\;.
\label{eq:theta_root}
\end{equation}
(The maximum angle is $\theta_{\rm max} = \pi - \theta_{\rm min}$.)  A
useful reparameterization of the $\theta$ coordinate is
\begin{equation}
\cos\theta = \cos\theta_{\rm min}\cos\chi\;,
\label{eq:theta_of_chi}
\end{equation}
where $\chi$ accumulates like the angle $\psi_r$.  By substitution in
Eq.\ (\ref{eq:theta_of_tau}), we can easily develop an equation
governing the evolution of $\chi$.

Before moving on, we note that Eqs.\ (\ref{eq:radial_turning_points})
and (\ref{eq:theta_root}) allow us to map from the parameters $(E,
L_z, Q)$ to $(p,e,\theta_{\rm min})$.  [Schmidt {\cite{schmidt}} in
fact gives an analytic solution for ($E, L_z, Q$) as functions of $(p,
e, \theta_{\rm min})$.]  Up to initial conditions, either
parameterization thus completely specifies an orbit.  We will flip
between these parameterizations as convenient.

\subsection{Orbital frequencies for black holes}
\label{sec:normal_bh_freqs}

To frame our discussion, we begin by examining the frequencies of
motion for ``normal'' black hole spacetimes.  Our discussion here
closely follows that given by Schmidt {\cite{schmidt}}, which uses
Hamilton-Jacobi methods to compute black hole orbital frequencies.  A
useful starting point is to note that in separating the
coordinate-space motion, one not only identifies constants of the
motion, but also the action variables
\begin{eqnarray}
J_r &\equiv& \frac{1}{2\pi} \oint p_r\,dr =
\frac{1}{\pi} \int_{r_p}^{r_a}\frac{\sqrt{R(r)}}{\Delta}\,dr\;,
\label{eq:Jr} \\
J_\theta &\equiv& \frac{1}{2\pi} \oint p_\theta\,d\theta =
\frac{2}{\pi}\int_{\theta_{\rm
min}}^{\pi/2}\sqrt{\Theta(\theta)}\,d\theta\;,
\label{eq:Jtheta} \\
J_\phi &\equiv& \frac{1}{2\pi} \oint p_\phi\,d\phi = L_z \;.
\label{eq:Jphi}
\end{eqnarray}
It is also useful to define
\begin{equation}
J_t \equiv -E\;.
\label{eq:Jt}
\end{equation}
This is a slight abuse of the notation since geodesic motion is not
cyclic in $t$ (and hence we cannot define $J_t$ as a closed integral
over time), but is convenient for reasons we will illustrate shortly.

At least formally, we can now reparameterize our Hamiltonian
(\ref{eq:Hamiltonian_def}) in terms of the action variables $J_\mu$.
Let us write the Hamiltonian so reparameterized as
$\Hcal^{\rm{(aa)}}$.  By Hamilton-Jacobi theory, the orbital
frequencies are the derivatives of $\Hcal^{\rm{(aa)}}$ with respect to
the action variables:
\begin{equation}
m\omega^i = \pder{\Hcal^{(\mathrm{aa})}}{J_i}\;.
\label{eq:freqs_def}
\end{equation}
For black hole orbits, we cannot explicitly reparameterize the
Hamiltonian in this way, making it difficult to calculate the orbital
frequencies.  However, by using the chain rule, we can re-express
(\ref{eq:freqs_def}) in terms of derivatives that are not so difficult
to explicitly write out.  Following Ref.\ {\cite{schmidt}} (modifying
its notation slightly), we put
\begin{equation}
P_\beta \doteq (\Hcal, E, L_z, Q)\;.
\label{eq:P_def}
\end{equation}
Define the matrices ${\cal A}$ and ${\cal B}$, whose components are
\begin{equation}
{{\cal A}_\alpha}^\beta = \frac{\partial P_\alpha}{\partial
J_\beta}\;, \qquad {{\cal B}_\alpha}^\beta = \frac{\partial
J_\alpha}{\partial P_\beta}\;.
\label{eq:A_and_B}
\end{equation}
By the chain rule, these matrices have an inverse relationship:
\begin{equation}
{{\cal A}_\alpha}^\beta{{\cal B}_\beta}^\gamma =
{\delta_\alpha}^\gamma\;.
\label{eq:chain}
\end{equation}
The components of the matrix ${\cal A}$ are directly related to the
frequencies we wish to compute.  In particular, since $P_t$ is just
the invariant Hamiltonian, $m\omega^i = \partial P_t/\partial J_i
\equiv {{\cal A}_t}^i$.  However, the components of the matrix ${\cal
B}$ are written in a way that is fairly easy to work out.  We exploit
this to write $m\omega^i = {({\cal B}^{-1})_t}^i$, from which we find
\begin{eqnarray}
m\omega^r &=& \frac{\partial J_\theta/\partial Q}{(\partial
J_r/\partial\Hcal)(\partial J_\theta/\partial Q) - (\partial
J_r/\partial Q)(\partial J_\theta/\partial\Hcal)} \;,
\label{Kerr_exact_omegar}
\\
m\omega^\theta &=& \frac{-\partial J_r/\partial Q}{(\partial
J_r/\partial\Hcal)(\partial J_\theta/\partial Q) - (\partial
J_r/\partial Q)(\partial J_\theta/\partial\Hcal)} \;,
\label{Kerr_exact_omegatheta}
\\
m\omega^\phi &=& \frac{(\partial J_r/\partial Q)(\partial
J_\theta/\partial L_z) - (\partial J_r/\partial L_z)(\partial
J_\theta/\partial Q)}{(\partial J_r/\partial\Hcal)(\partial
J_\theta/\partial Q) - (\partial J_r/\partial Q)(\partial
J_\theta/\partial\Hcal)} \;.
\label{Kerr_exact_omegaphi}
\end{eqnarray}
The partial derivatives of $J_r$ and $J_\theta$ appearing here are
given by
\begin{eqnarray}
\pder{J_r}{\Hcal} &=& \frac{1}{2\pi}\oint \frac{r^2}{\sqrt{R(r)}} dr
\;,
\\
\pder{J_r}{Q} &=& -\frac{1}{4\pi} \oint \frac{1}{\sqrt{R(r)}} dr \;,
\\
\pder{J_r}{L_z} &=& -\frac{1}{2\pi}\oint \frac{r(r L_z - 2M(L_z -
aE))}{\Delta\sqrt{R(r)}} dr \;,
\\
\pder{J_r}{E} &=& \frac{1}{2\pi}\oint \frac{r\left[r E(r^2 + a^2)-2 M
a(L_z - aE)\right]}{\Delta\sqrt{R(r)}} dr\;,
\\
\pder{J_\theta}{\Hcal} &=& \frac{1}{2\pi}\oint
\frac{a^2\cos^2\theta}{\sqrt{\Theta(\theta)}} d\theta \;,
\\
\pder{J_\theta}{Q} &=&  \frac{1}{4\pi} \oint
\frac{1}{\sqrt{\Theta(\theta)}} d\theta \;,
\\
\pder{J_\theta}{L_z} &=& -\frac{1}{2\pi}\oint
\frac{L_z\cot^2\theta}{\sqrt{\Theta(\theta)}} d\theta \;,
\\
\pder{J_\theta}{E} &=& \frac{1}{2\pi}\oint \frac{a^2
E\cos^2\theta}{\sqrt{\Theta(\theta)}}d\theta\;.
\end{eqnarray}
(The derivatives $\partial J_{r,\theta}/\partial E$ will be needed for
a quantity we introduce below.)  Schmidt {\cite{schmidt}} combines
these results to give closed-form expressions for the three
frequencies $\omega^{r, \theta, \phi}$; we will not repeat these
expressions here.

The frequencies $\omega^{r,\theta,\phi}$ are conjugate to the orbit's
{\it proper} time; they would be measured by an observer who rides on
the orbit itself.  For our purposes, it will be more useful to convert
to frequencies conjugate to the Boyer-Lindquist coordinate time,
describing measurements made by a distant observer.  The
quantity\footnote{We have adjusted notation from Schmidt slightly to
avoid confusion with our metric function $\gamma$.}
\begin{equation}
\Gamma \equiv \frac{1}{m}\frac{\partial\Hcal^{\rm (aa)}}{\partial J_t} =
-\frac{1}{m}\frac{\partial\Hcal^{\rm (aa)}}{\partial E}
\end{equation}
performs this conversion; the frequencies $\Omega^{r,\theta,\phi} =
\omega^{r,\theta,\phi}/\Gamma$ are of observational relevance.  Going
back to Eqs.\ (\ref{eq:A_and_B}) and (\ref{eq:chain}), we find
$m\Gamma = {({\cal B}^{-1})_t}^t$, or
\begin{equation}
m\Gamma = \frac{(\partial J_r/\partial E)(\partial J_\theta/\partial
Q) - (\partial J_r/\partial Q)(\partial J_\theta/\partial
E)}{(\partial J_r/\partial\Hcal)(\partial J_\theta/\partial Q) -
(\partial J_r/\partial Q)(\partial J_\theta/\partial\Hcal)} \;.
\label{Kerr_exact_Gamma}
\end{equation}

In the discussion that follows, it will be useful to have weak-field
($p \gg M$) forms of these frequencies as a point of comparison.  We
begin by we taking the exact expressions for $E$, $L_z$, and $Q$ given
by Schmidt (Appendix B of Ref.\ {\cite{schmidt}}), expand to leading
order in $a$, and then expand in $1/p$.  The result is
\begin{eqnarray}
E &=& m\left[1 - \frac{1 - e^2}{2p} + \frac{(1 - e^2)^2}{p^2}
\left(\frac{3}{8} - \frac{a}{M}\frac{\sin\theta_{\rm
min}}{\sqrt{p}}\right)\right]\;,
\label{eq:Eweak_Kerr}\\
L_z^2 + Q &=& m^2M^2p\left[1 + \frac{3 + e^2}{p} + \frac{(3 +
e^2)^2}{p^2} - \frac{a}{M}\left(\frac{2(3 + e^2)}{p^{3/2}} +
\frac{4(2+e^2)(3+e^2)}{p^{5/2}}\right)\sin\theta_{\rm min}\right]\;,
\label{eq:QplusLzSqrweak_Kerr}\\
\frac{L_z}{\sqrt{L_z^2 + Q}} &=& \sin\theta_{\rm min}\;.
\label{eq:sinthetamweak_Kerr}
\end{eqnarray}
The orbital frequencies become
\begin{eqnarray}
\Omega^r &=& \omega^K \left[ 1 - \frac{3(1-e^2)}{p} +
\frac{a}{M}\frac{3(1-e^2)\sin\theta_{\rm min}}{p^{3/2}}\right] \:,
\\
\Omega^\theta &=& \omega^K \left[ 1 + \frac{3e^2}{p} -
\frac{a}{M}\frac{3(1+e^2)\sin\theta_{\rm min}}{p^{3/2}}\right] \:,
\\
\Omega^\phi &=& \omega^K \left[ 1 + \frac{3e^2}{p} -
\frac{a}{M}\frac{3(1+e^2)\sin\theta_{\rm min}}{p^{3/2}}
+ \frac{a}{M}\frac{2}{p^{3/2}}\right] \:,
\end{eqnarray}
where
\begin{equation}
\omega^K = \frac{1}{M} \left(\frac{1 - e^2}{p}\right)^{3/2}
\end{equation}
is the Kepler frequency.

\subsection{Orbital frequencies of bumpy black holes}
\label{sec:bumpy_bh_freqs}

Now examine how the orbital frequencies change if the black hole is
bumpy.  Begin with the Hamiltonian ${\cal H}$.  It remains conserved
with value $-m^2/2$, but its functional form is shifted:
\begin{eqnarray}
\Hcal &=& \frac{1}{2}g^{\alpha\beta} p_\alpha p_\beta = -\frac{m^2}{2}
\nonumber\\
&=& \hat\Hcal + \Hcal_1\;,
\label{eq:bumpy_hamiltonian}
\end{eqnarray}
where $\hat\Hcal$ is the original (non-bumpy) Hamiltonian and
$\Hcal_1$ gathers together the influence of the spacetime's bumpiness.
To first order in $b_{\alpha\beta}$,
\begin{equation}
g^{\alpha\beta} = {\hat g}^{\alpha\beta} - b^{\alpha\beta}\;,
\end{equation}
where
\begin{equation}
b^{\alpha\beta} = {\hat g}^{\alpha\mu}{\hat g}^{\beta\nu}b_{\mu\nu}\;.
\end{equation}
Combining these expressions, we have
\begin{equation}
\Hcal_1 = -\frac{1}{2}{\hat g}^{\alpha\mu}{\hat
g}^{\beta\nu}b_{\mu\nu} p_\alpha p_\beta\;,
\end{equation}
which can be rewritten
\begin{equation}
\Hcal_1 = -\frac{1}{2}b_{\mu\nu} p^\mu p^\nu =
-\frac{m^2}{2}b_{\mu\nu} \frac{dx^\mu}{d\tau} \frac{dx^\nu}{d\tau}\;.
\end{equation}

When we add bumps to a spacetime, shifting the Hamiltonian by
$\Hcal_1$, the motion is no longer separable (except for the
special case of equatorial motion) and the techniques used in
Sec.\ {\ref{sec:normal_bh_freqs}} for computing orbital frequencies do
not work.  However, since the spacetime is ``close to'' the exact
black hole spacetime in a well-defined sense, the motion is likewise
``close to'' the integrable motion.  We can thus take advantage of
canonical perturbation theory as described in, for example, Goldstein
{\cite{goldstein}}, to calculate how the spacetime's bumps change the
frequencies.

The key result which we use is that the shift can be found by suitably
{\it averaging} $\Hcal_1$:
\begin{equation}
m\,\delta\omega^i = \frac{\partial
\langle\Hcal_1\rangle}{\partial{\hat J}_i}\;,\qquad m\,\delta\Gamma
= \frac{\partial \langle\Hcal_1\rangle}{\partial{\hat J}_t}\;.
\label{eq:gen_shifts}
\end{equation}
Notice that the derivatives are taken with respect to the action
variables defined for the background motion; the averaging, which we
denote with angular brackets, is likewise
done with respect to orbits in the background.  Once these derivatives
are taken, it is simple to compute the changes to the observable
frequencies. Expanding
\begin{eqnarray}
\Omega^i &=& \frac{\omega^i}{\Gamma} =
\frac{\hat\omega^i + \delta\omega^i}{\hat\Gamma + \delta\Gamma}
\nonumber\\
&\equiv& \hat\Omega^i + \delta\Omega^i\;,
\end{eqnarray}
we read out
\begin{equation}
\delta\Omega^i = \frac{\delta\omega^i}{\hat\Gamma} -
\frac{\hat\omega^i\,\delta\Gamma}{\hat\Gamma^2}\;,
\end{equation}
or
\begin{equation}
\frac{\delta\Omega^i}{\Omega^i} = \frac{\delta\omega^i}{\hat\omega^i}
- \frac{\delta\Gamma}{\hat\Gamma}\;.
\label{eq:freq_shift}
\end{equation}

The averaging used in Eq.\ (\ref{eq:gen_shifts}) in described in
detail in Appendix {\ref{app:averaging}}.  This procedure uses the
fact that the radial and polar components of the background motion can
be separated, and thus can be averaged independently.  Following
Appendix {\ref{app:averaging}}, this amounts to computing
\begin{equation}
\langle\Hcal_1\rangle = \frac{1}{\Upsilon^t(2\pi)^2}\int_0^{2\pi}
dw^r \int_0^{2\pi} dw^\theta\,\Hcal_1\left[r(w^r),
\theta(w^\theta)\right] T\left[r(w^r),\theta(w^\theta)\right] \;,
\label{eq:H1_averaged}
\end{equation}
where $T(r,\theta)$ is defined by Eq.\ (\ref{eq:t_of_tau}),
$\Upsilon^t$ is defined by Eq.\ (\ref{eq:Upsilon_t}), and where
$w^{r,\theta}$ are angles associated with the separated $r$ and
$\theta$ motions (defined and discussed in detail in Appendix
{\ref{app:averaging}}).

Before moving on, we discuss a few issues in practically computing
$\langle\Hcal_1\rangle$ and the frequency shifts.  Begin by expanding
the Hamiltonian:
\begin{eqnarray}
{\cal H}_1 &=& -\frac{m^2}{2}\left[b_{tt}\left(\frac{dt}{d\tau}\right)^2
+ b_{rr}\left(\frac{dr}{d\tau}\right)^2
+ b_{\theta\theta}\left(\frac{d\theta}{d\tau}\right)^2
+ b_{\phi\phi}\left(\frac{d\phi}{d\tau}\right)^2
+ 2b_{tr}\frac{dt}{d\tau}\frac{dr}{d\tau}
+ 2b_{\phi r}\frac{d\phi}{d\tau}\frac{dr}{d\tau}
+ 2b_{\phi t}\frac{d\phi}{d\tau}\frac{dt}{d\tau}\right]
\nonumber\\
&=& -\frac{m^2}{2}\left[b_{tt}\left(\frac{dt}{d\tau}\right)^2
+ b_{rr}\left(\frac{dr}{d\tau}\right)^2
+ b_{\theta\theta}\left(\frac{d\theta}{d\tau}\right)^2
+ b_{\phi\phi}\left(\frac{d\phi}{d\tau}\right)^2
+ 2b_{\phi t}\frac{d\phi}{d\tau}\frac{dt}{d\tau}\right]
\nonumber\\
&=& -\frac{1}{2\Sigma^2}\left[b_{tt}T(r,\theta)^2
+ b_{rr}R(r) + b_{\theta\theta}\Theta(\theta)
+ b_{\phi\phi}\Phi(r,\theta)^2
+ 2b_{\phi t}\Phi(r,\theta)T(r,\theta)\right]\;.
\label{eq:H_for_averaging}
\end{eqnarray}
In going from the first line to the second, we use the fact that terms
linear in $dr/d\tau$ go to zero when we average since the radial
motion switches sign after half a cycle.  The final line of Eq.\
(\ref{eq:H_for_averaging}) is in a good form for averaging.

In computing this average, we end up with $\langle\Hcal_1\rangle$ as a
function of $p$, $e$, and $\theta_{\rm min}$.  We likewise compute the
actions $J_\mu$ using these parameters, and then compute the shifts to
the frequencies and $\Gamma$ using the chain rule.  To set up this
calculation, define an array $b_\beta$ which contains all the system's
physical parameters:
\begin{equation}
b_\beta \doteq \left(m, p, e, \theta_{\rm min}\right)\;.
\label{eq:b_def}
\end{equation}
Next, define the matrix ${\cal J}$, the Jacobian of the actions with
respect to these parameters:
\begin{equation}
{({\cal J})_\alpha}^\beta = \frac{\partial J_\alpha}{\partial b_\beta}
\end{equation}
Then,
\begin{eqnarray}
\delta\omega^i &=& \frac{\partial{\langle\Hcal_1\rangle}}{\partial
b_\alpha}{({\cal J}^{-1})_\alpha}^i\;,
\label{eq:delta_omega}\\
\delta\Gamma &=& \frac{\partial{\langle\Hcal_1\rangle}}{\partial
b_\alpha}{({\cal J}^{-1})_\alpha}^t\;,
\label{eq:delta_Gamma}
\end{eqnarray}
where ${\cal J}^{-1}$ is the matrix inverse of the Jacobian ${\cal
J}$.

\section{Results I: Orbits of bumpy Schwarzschild black holes}
\label{sec:results_schw}

We now examine spacetimes and orbits for specific choices of $\psi_1$.
Recalling that this function satisfies the Laplace equation, we take
$\psi_1$ to be a pure multipole in the ``Weyl sector'' [i.e., in the
coordinates of Eq.\ (\ref{eq:weyl})].  As we will show in this section
and the next, this smoothes out the bumps and cures the strong-field
pathologies associated with orbits in the spacetimes developed in
CH04.

Note that a pure multipole in the Weyl sector will {\it not}
correspond to a pure Geroch-Hansen moment of the black hole.  For
example, taking $\psi_1$ to be proportional to an $l = 2$ spherical
harmonic does not change only the moment $M_2$ of the Geroch-Hansen
sequence [Eq.\ (\ref{eq:Kerr_moments})].  However, it turns out that
taking $\psi_1$ to be proportional to a spherical harmonic $Y_{l0}$
changes no Geroch-Hansen moments {\it lower} than $M_l$: taking
$\psi_1$ to be an $l = 2$ harmonic changes $M_2$ and higher moments;
taking it to be an $l = 3$ harmonic changes $M_3$ and higher; etc.  A
companion paper {\cite{vigeland}} demonstrates this explicitly, and
further shows that the dominant change for $\psi_1 \propto Y_{l0}$ is
to the $l$-th Geroch-Hansen moment.  Further, since the equations
governing $\psi_1$ and $\gamma_1$ are linear in these fields, one can
choose $\psi_1$ to be a combination of multipoles such that the
resulting spacetime puts its ``bump'' into a single Geroch-Hansen
moment.  In this way, one can arbitrarily adjust the Geroch-Hansen
moments of a spacetime (providing that the adjustments are small).

We begin with spacetimes and orbits of bumpy Schwarzschild black
holes.

\subsection{Quadrupole bumps ($l=2$)}

First we examine an $l = 2$ perturbation in the Weyl sector.  The
perturbation $\psi_1$ which satisfies Eq.\ (\ref{eq:weyl_einstein1})
and has an $l = 2$ spherical harmonic form is
\begin{eqnarray}
\psi_1^{l=2}(\rho,z) = B_2M^3\frac{Y_{20}(\theta_{\rm Weyl})}{(\rho^2 +
z^2)^{3/2}} = \frac{B_2M^3}{4}\sqrt{\frac{5}{\pi}}
\frac{3\cos^2\theta_\mathrm{Weyl} - 1}{(\rho^2 + z^2)^{3/2}}
\label{l2_potential}
\end{eqnarray}
where $\cos\theta_{\rm Weyl} = z/\sqrt{\rho^2 + z^2}$.  The
dimensionless constant $B_2$ sets the magnitude of the spacetime's
bumpiness for this multipole.  Since we are treating the bumpiness as
a perturbation, $B_2 \ll 1$.  Transforming to Schwarzschild
coordinates by Eqs.\ (\ref{eq:rho_of_uv}), (\ref{eq:z_of_uv}),
(\ref{eq:r_of_u}), and (\ref{eq:theta_of_v}), we find
\begin{eqnarray}
\psi_1^{l = 2}(r,\theta) &=& \frac{B_2M^3}{4}\sqrt{\frac{5}{\pi}}
\frac{1}{d(r,\theta)^3} \left[\frac{3(r -
M)^2\cos^2\theta}{d(r,\theta)^2} - 1\right]\;,
\end{eqnarray}
where
\begin{equation}
d(r,\theta) \equiv (r^2 - 2Mr + M^2\cos^2\theta)^{1/2}\;.
\end{equation}
As a useful aside, the mapping from $(\rho,z)$ to $(r,\theta)$ implies
that any Weyl sector $\psi_1$ can be transformed into Schwarzschild
coordinates by putting
\begin{eqnarray}
\rho^2 + z^2 &\to& d(r,\theta)\;,
\label{eq:rw_map}
\\
\cos\theta_{\rm Weyl} &\to& \frac{(r-M)}{d(r,\theta)}\cos\theta\;.
\label{eq:thetaw_map}
\end{eqnarray}

Integrating the constraint (\ref{gamma_constraint}) and imposing the
condition $\gamma_1(r\to\infty) = 0$ gives
\begin{equation}
\gamma_1^{l = 2}(r,\theta) = B_2\sqrt\frac{5}{\pi}\left[ \frac{(r -
M)}{2}\frac{\left[c_{20}(r) + c_{22}(r)\cos^2\theta\right]}{d(r,\theta)^5}
- 1\right]\;,
\end{equation}
where
\begin{eqnarray}
c_{20}(r) &=& 2(r-M)^4 - 5M^2(r-M)^2 + 3M^4\;,
\label{eq:g20_schw}\\
c_{22}(r) &=& 5M^2(r-M)^2 - 3M^4\;.
\label{eq:g22_schw}
\end{eqnarray}

\begin{figure}[ht]
\includegraphics[width=5.8cm]{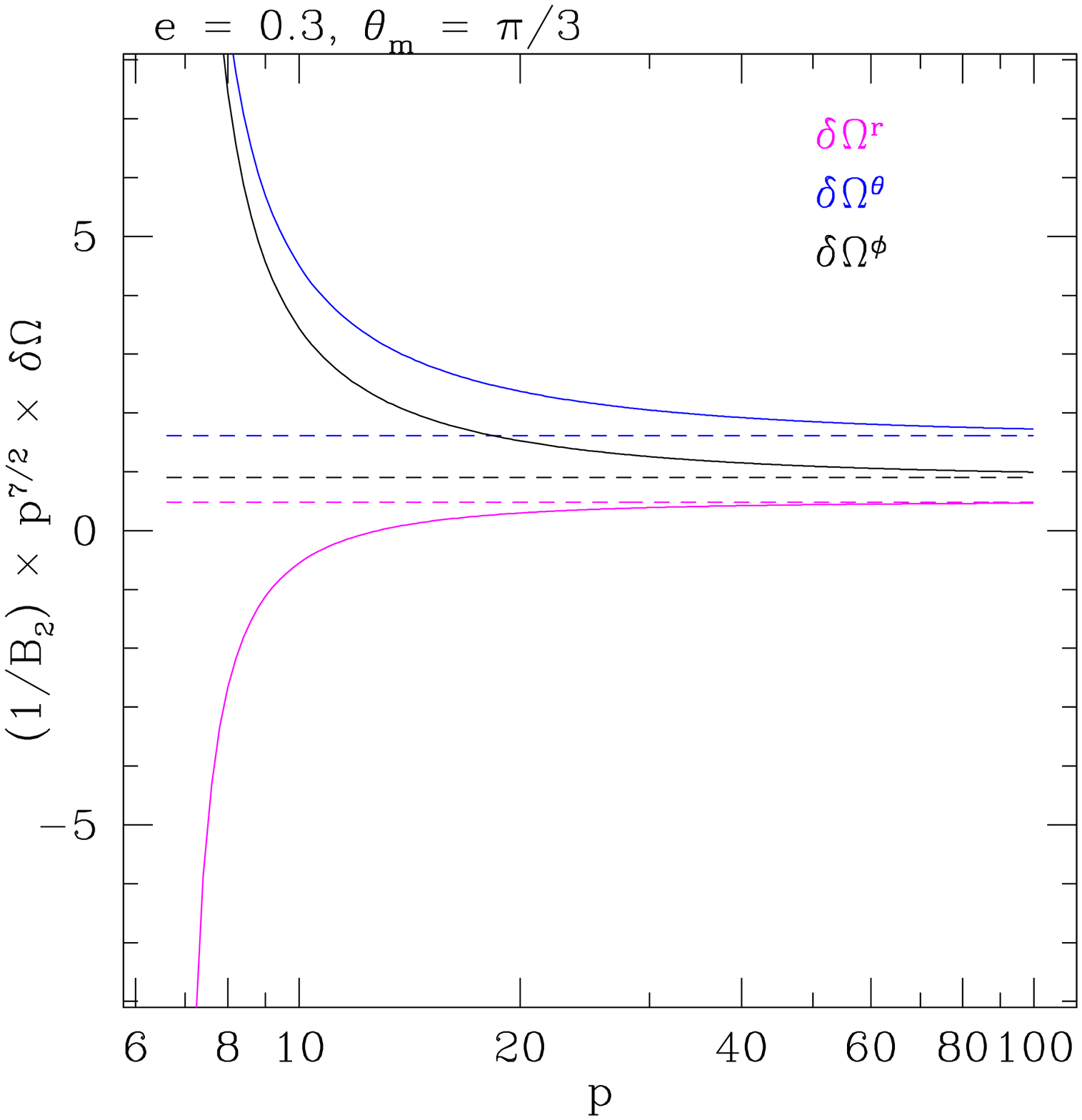}
\includegraphics[width=5.8cm]{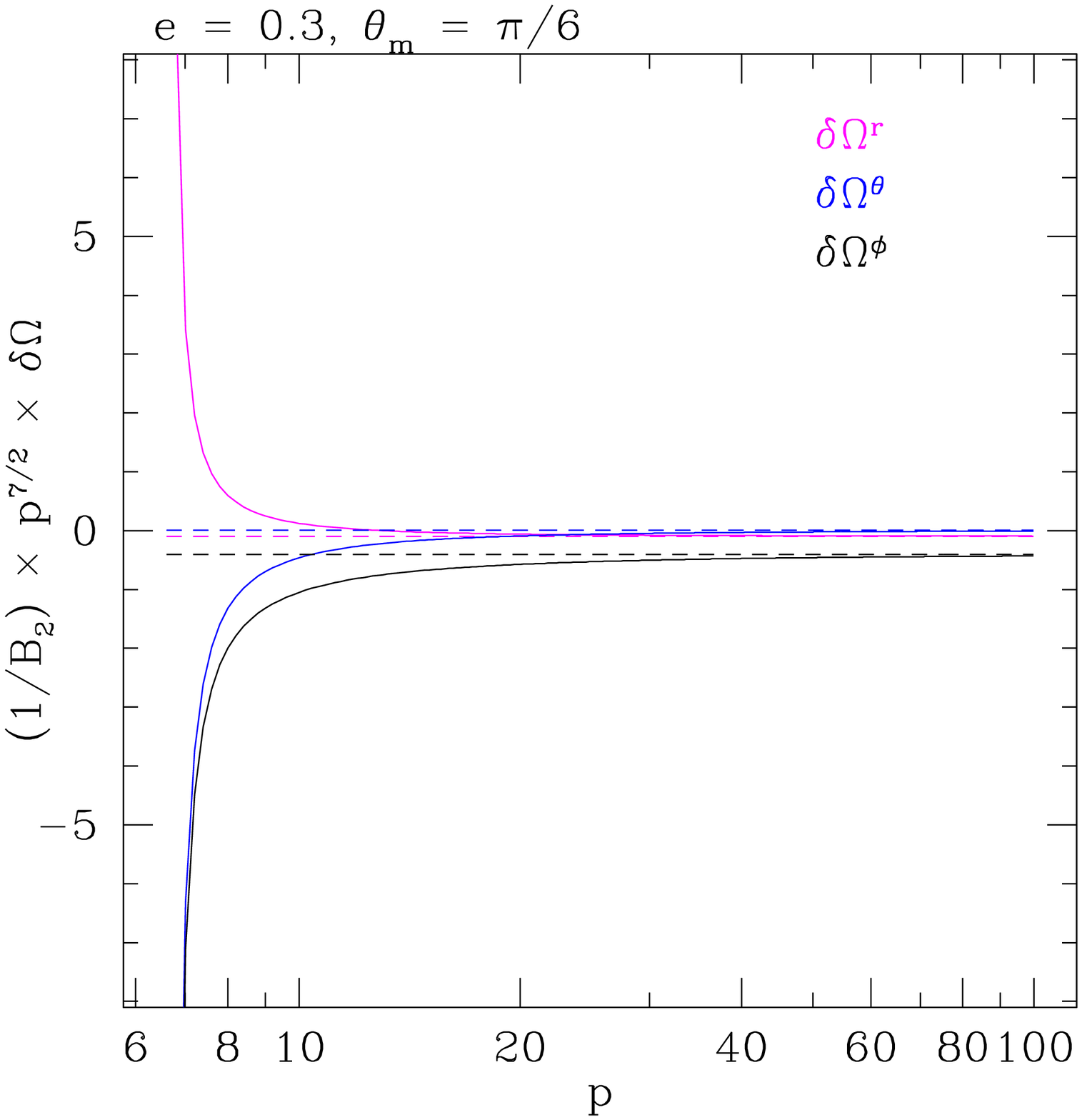}
\includegraphics[width=5.8cm]{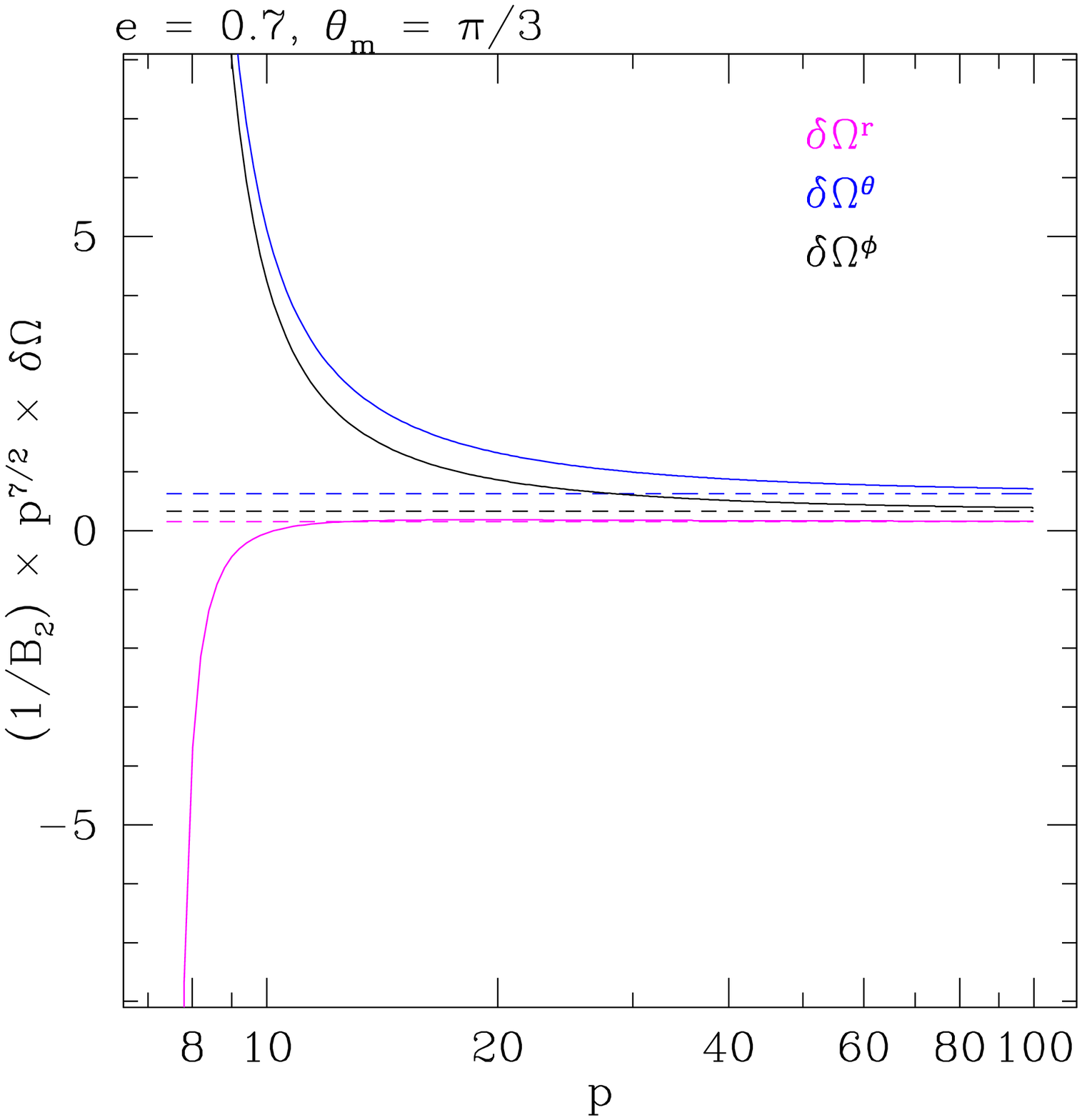}
\caption{Shifts to black hole orbital frequencies due to an $l = 2$
bump.  The shifts $\delta\Omega^{r,\theta,\phi}$ are normalized by the
bumpiness parameter $B_2$, and are scaled by $p^{7/2}$; this is
because in the Newtonian limit, $\delta\Omega^{r,\theta,\phi} \propto
p^{-7/2}$.  The Newtonian limit (dashed lines) does a good describing
the exact calculations (solid lines) for large $p$.  This limit
substantially underestimates the shifts in the strong field.  Notice
that the radial frequency shift changes sign in the strong field,
typically at $p \sim (10 - 13)M$, depending slightly on parameters.
This behavior is starkly different from the weak field limit.}
\label{fig:dOm_l2}
\end{figure}

Figure {\ref{fig:dOm_l2}} shows the impact of an $l = 2$ bump on
orbital frequencies as a function of $p$ for a few choices of $e$ and
$\theta_m$.  We show the three shifts $\delta\Omega^{r,\theta,\phi}$,
normalized by the bumpiness $B_2$ and rescaled by the asympotic
weak-field dependence $\delta\Omega^x_{l = 2} \propto p^{-7/2}$
(derived in Appendix {\ref{app:newton}}).  As we move to the weak
field, the numerical results (solid lines) converge to the weak-field
forms (dashed lines).  The frequency shifts generically get
substantially larger as we move into the strong field.  The bumps have
a very strong influence near the last stable orbit, $p_{\rm LSO} =
(6+2e)$, although the behavior is smooth and non-pathological.

\subsection{Octupole bumps ($l=3$)}

Next, consider an $l = 3$ perturbation.  In Weyl coordinates, we put
\begin{equation}
\psi_1^{l = 3}(\rho,z) = B_3M^4 \frac{Y_{30}(\theta_{\rm Weyl})}{(\rho^2 +
z^2)^2}\;;
\label{l3_potential}
\end{equation}
In Schwarzschild coordinates, this becomes
\begin{equation}
\psi_1^{l = 3}(r,\theta) = \frac{B_3M^4}{4}\frac{1}{d(r,\theta)^4}
\sqrt{\frac{7}{\pi}} \left[\frac{5(r -
M)^3\cos^3\theta}{d(r,\theta)^3} - \frac{3(r -
M)\cos\theta}{d(r,\theta)}\right] \;.
\end{equation}
From the constraint equation (\ref{gamma_constraint}) and the
condition $\gamma_1(r\to\infty)=0$, we find
\begin{equation}
\gamma_1^{l = 3}(r,\theta) = \frac{B_3 M^5}{2}
\sqrt{\frac{7}{\pi}} \cos\theta \left[\frac{c_{30}(r) +
c_{32}(r)\cos^2\theta + c_{34}(r)\cos^4\theta +
c_{36}(r)\cos^6\theta}{d(r,\theta)^7}\right] \;,
\end{equation}
where
\begin{eqnarray}
c_{30}(r) &=& -3r(r - 2M)\;,
\label{eq:g30_schw}\\
c_{32}(r) &=& 10r(r - 2M) + 2M^2\;,
\label{eq:g32_schw}\\
c_{34}(r) &=& -7r(r - 2M)\;,
\label{eq:g34_schw}\\
c_{36}(r) &=& -2M^2\;.
\label{eq:g36_schw}
\end{eqnarray}

Notice that $\psi_1^{l=3}$ and $\gamma_1^{l=3}$ are proportional to
$\cos\theta$.  As such, their contribution to the averaged Hamiltonian
$\langle\Hcal_1\rangle$ is {\it zero}: There is {\it no} secular shift
to orbital frequencies from $l = 3$ bumps.  This is identical to the
result in Newtonian gravity, as discussed in Appendix
{\ref{app:newton}}, and holds for all odd values of $l$.

As in the Newtonian limit, there will be {\it non}-secular shifts to
the motion which cannot be described by our orbit-averaged approach.
These shifts would be apparent in a direct (time-domain) evolution of
the geodesics of spacetimes with odd $l$ bumps.  It would be a useful
exercise to examine these effects and ascertain under which conditions
odd $l$ spacetime bumps could, in principle, have an observable
impact.

\subsection{Hexadecapole bumps ($l=4$)}

We conclude our discussion of Schwarzschild bumps with $l = 4$:
\begin{equation}
\psi_1^{l = 4}(\rho,z) = B_4M^5 \frac{Y_{40}(\theta_{\rm Weyl})}{(\rho^2 +
z^2)^{5/2}}\;,
\label{l4_potential}
\end{equation}
from which we obtain
\begin{equation}
\psi_1^{l = 4}(r,\theta) = \frac{B_4M^5}{16}\frac{1}{d(r,\theta)^5}
\sqrt{\frac{9}{\pi}} \left[\frac{35(r -
M)^4\cos^4\theta}{d(r,\theta)^4} - \frac{30(r -
M)^2\cos^2\theta}{d(r,\theta)^2} + 3\right]\;.
\end{equation}
Solving for $\gamma_1$ as before, we find
\begin{equation}
\gamma_1^{l = 4}(r,\theta) = B_4\sqrt{\frac{9}{\pi}}
\left[\frac{(r-M)}{2}\frac{c_{40}(r) + c_{42}(r)\cos^2\theta +
c_{44}(r)\cos^4\theta}{d(r,\theta)^9} - 1\right] \;,
\end{equation}
where
\begin{eqnarray}
c_{40}(r) &=& 8(r-M)^8 - 36M^2(r-M)^6 + 63M^4(r-M)^4 - 50M^6(r-M)^2 +
15M^8\;,
\label{eq:g40_schw}\\
c_{42}(r) &=& 36M^2(r-M)^6 - 126M^4(r-M)^4 + 120M^6(r-M)^2 - 30M^8\;,
\label{eq:g42_schw}\\
c_{44}(r) &=& 63M^4(r-M)^4 - 70M^6(r-M)^2 + 15M^8\;.
\label{eq:g44_schw}
\end{eqnarray}

\begin{figure}[ht]
\includegraphics[width=5.8cm]{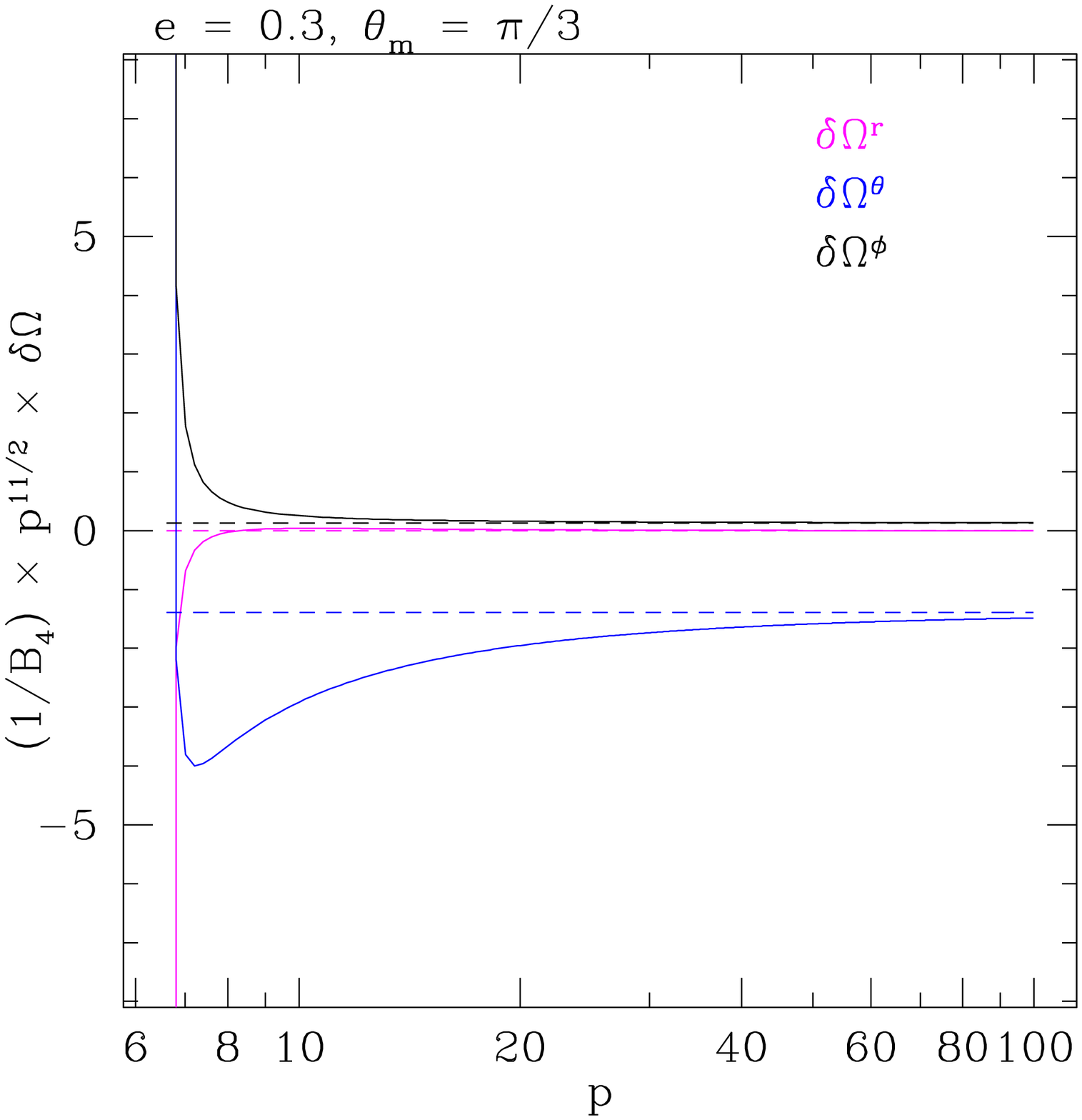}
\includegraphics[width=5.8cm]{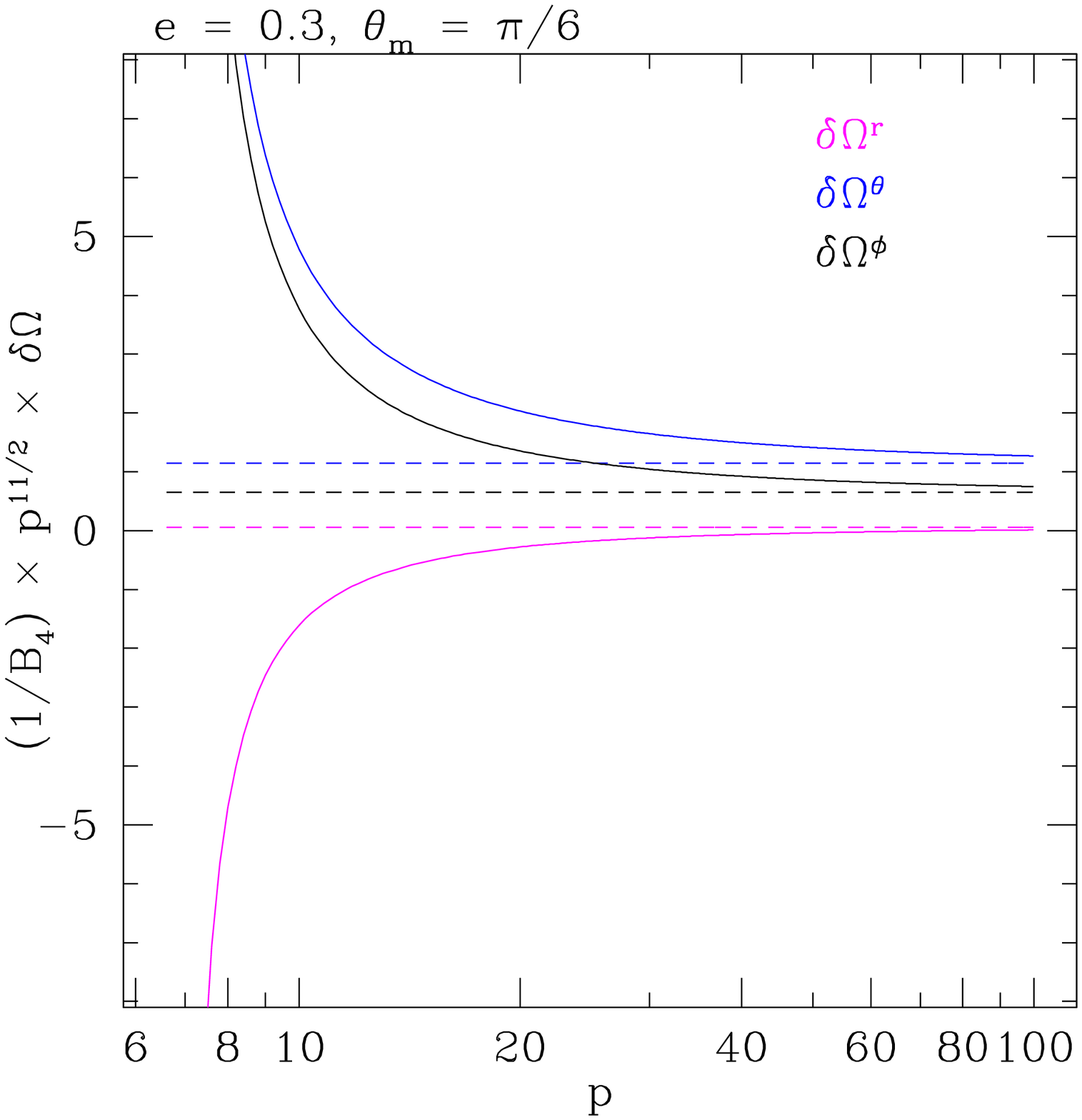}
\includegraphics[width=5.8cm]{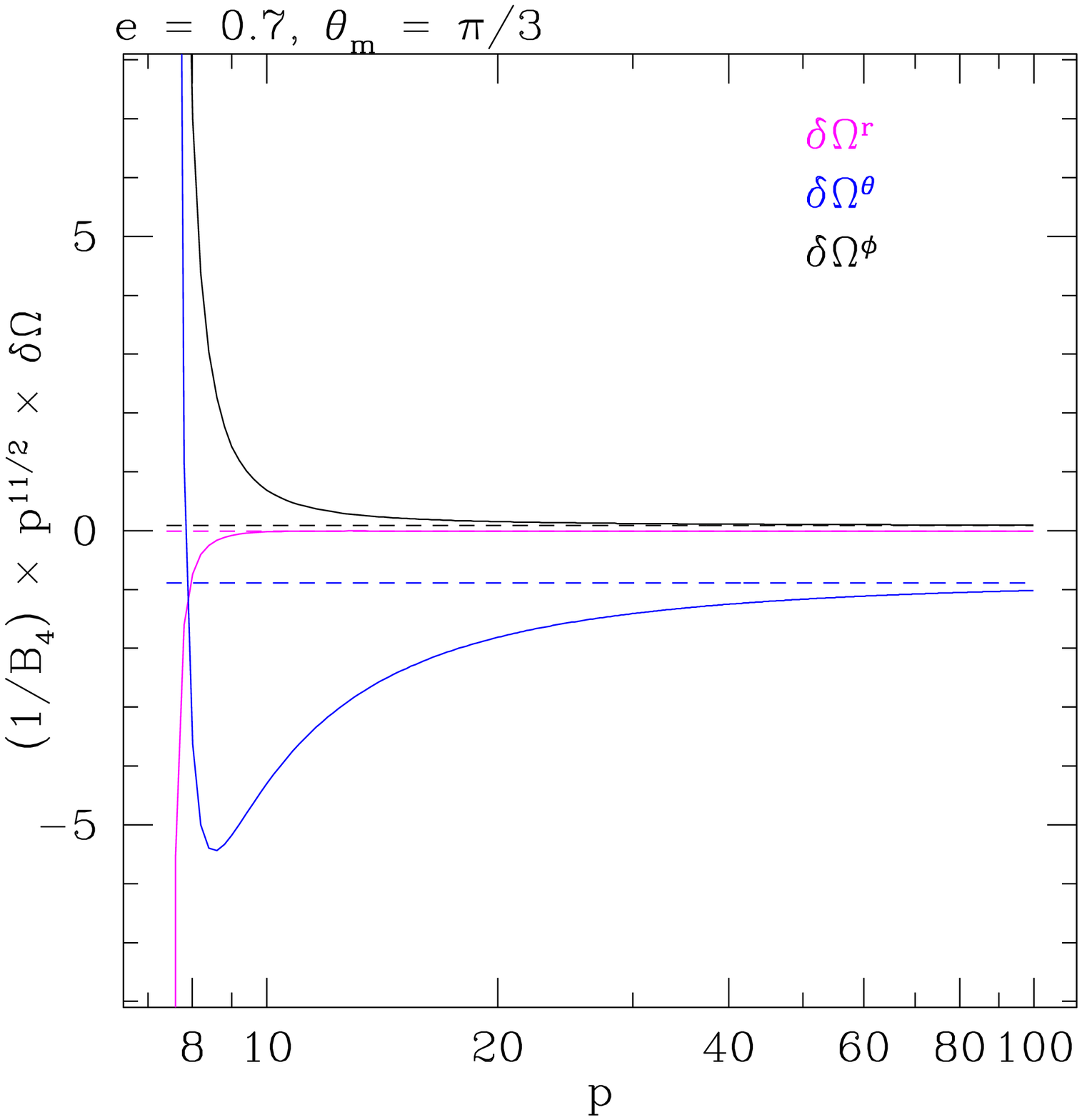}
\caption{Shifts to black hole orbital frequencies due to an $l = 4$
bump.  The shifts $\delta\Omega^{r,\theta,\phi}$ are normalized by the
bumpiness parameter $B_4$, and are scaled by $p^{11/2}$, which sets
the scaling in the Newtonian limit.  As in the $l = 2$ case (Fig.\
\ref{fig:dOm_l2}), exact results and the Newtonian limit coincide at
large $p$, but there are significant differences in the strong field.
The functional behavior of the radial frequency shift can be
especially complicated in this case.}
\label{fig:dOm_l4}
\end{figure}

Figure {\ref{fig:dOm_l4}} presents the same orbits as are shown in
Fig.\ {\ref{fig:dOm_l2}} for $l = 4$.  The shifts are normalized by
the bumpiness $B_4$ and rescaled by the weak-field form
$\delta\Omega^x_{l = 4} \propto p^{-11/2}$.  The qualitative behavior
is largely the same as for the quadrupole bump.  In particular, we see
once again that there are no strong-field pathologies in the orbit
shifts, and that the degree of shift due to spacetime bumps is
especially strong near the last stable orbit.  The strong-field radial
oscillations in $\delta\Omega^r$ are even more pronounced than they
were in the $l = 2$ case.  This appears to be a robust signature of
non-Kerr multipoles in the strong-field.

\section{Results II: Orbits of bumpy Kerr black holes}
\label{sec:results_kerr}

We now repeat the above exercises on a Kerr black hole background.

\subsection{Quadrupole bumps ($l=2$)}
\label{Kerr_l2}

We begin with an $l = 2$ perturbation in the Weyl sector, starting
with Eq.\ (\ref{l2_potential}).  Transforming to prolate spheroidal
coordinates by Eqs.\ (\ref{eq:rho_of_uv}) and (\ref{eq:z_of_uv}), this
becomes
\begin{eqnarray}
\psi^{l=2}_1(u,v) &=& \frac{B_2}{4}\sqrt{\frac{5}{\pi}}
\left(\frac{3\cosh^2u\cos^2v}{\sinh^2u \sin^2v + \cosh^2u\cos^2v} -
1\right) \left(\sinh^2u \sin^2v + \cosh^2u \cos^2v\right)^{-3/2} \;.
\end{eqnarray}
The corresponding $\gamma_1$ is
\begin{eqnarray}
\gamma^{l=2}_1(u,v) &=& {B_2}\sqrt{\frac{5}{\pi}}\left[ \frac{\cosh u
\left[4 - \cos 2v + (5\cos 2v - 1)\cosh 2u + \cosh
4u\right]}{8\left(\sinh^2u \sin^2v + \cosh^2u \cos^2v\right)^{5/2}} -
1\right] \;.
\end{eqnarray}
Following the logic of the Newman-Janis algorithm, we allow $u$ to be
complex, and replace $\cosh^2u$ with $\cosh u \cosh\bar{u}$ and
$\sinh^2u$ with $(\cosh u \cosh\bar{u}-1)$.  Making the coordinate
transformation
\begin{equation}
\cosh u = \frac{r - ia\cos\theta}{M} - 1 \;,\qquad
v = \theta
\end{equation}
puts the result in Boyer-Lindquist coordinates:
\begin{eqnarray}
\psi_1^{l=2}(r, \theta) &=& \frac{B_2M^3}{4}\sqrt{\frac{5}{\pi}}
\frac{1}{d(r,\theta,a)^3}\left[\frac{3L(r,\theta,a)^2\cos^2\theta}
{d(r,\theta,a)^2}- 1\right] \;,
\\
\gamma_1^{l=2}(r, \theta) &=& B_2\sqrt{\frac{5}{\pi}} \left[
\frac{L(r,\theta,a)}{2} \frac{\left[c_{20}(r,a) +
c_{22}(r,a)\cos^2\theta +
c_{24}(r,a)\cos^4\theta\right]}{d(r,\theta,a)^5} - 1\right]\;,
\end{eqnarray}
where
\begin{eqnarray}
d(r, \theta, a) &=& \sqrt{r^2 - 2Mr + (M^2 + a^2)\cos^2\theta}\;,
\\
L(r,\theta,a) &=& \sqrt{(r - M)^2 + a^2\cos^2\theta}\;,
\end{eqnarray}
and
\begin{eqnarray}
c_{20}(r,a) &=& 2(r-M)^4 - 5M^2(r-M)^2 + 3M^4\;,
\label{eq:g20_kerr}\\
c_{22}(r,a) &=& 5M^2(r-M)^2 - 3M^4 + a^2\left[4(r-M)^2 - 5M^2\right]
\label{eq:g22_kerr}\\
c_{24}(r,a) &=& a^2(2a^2 + 5M^2)\;.
\label{eq:g24_kerr}
\end{eqnarray}
Note that the result for $\psi_1$ can be found by taking the
Weyl-sector perturbation given by Eq.\ (\ref{l2_potential}) and
putting
\begin{eqnarray}
\rho^2 + z^2 &\to& d(r,\theta,a)\;,
\\
\cos\theta_{\rm Weyl} &\to&
\frac{L(r,\theta,a)}{d(r,\theta,a)}\cos\theta\;.
\end{eqnarray}
This is the Kerr analog to the mapping described in Eqs.\
(\ref{eq:rw_map}) and (\ref{eq:thetaw_map}).

\begin{figure}[ht]
\includegraphics[width=5.8cm]{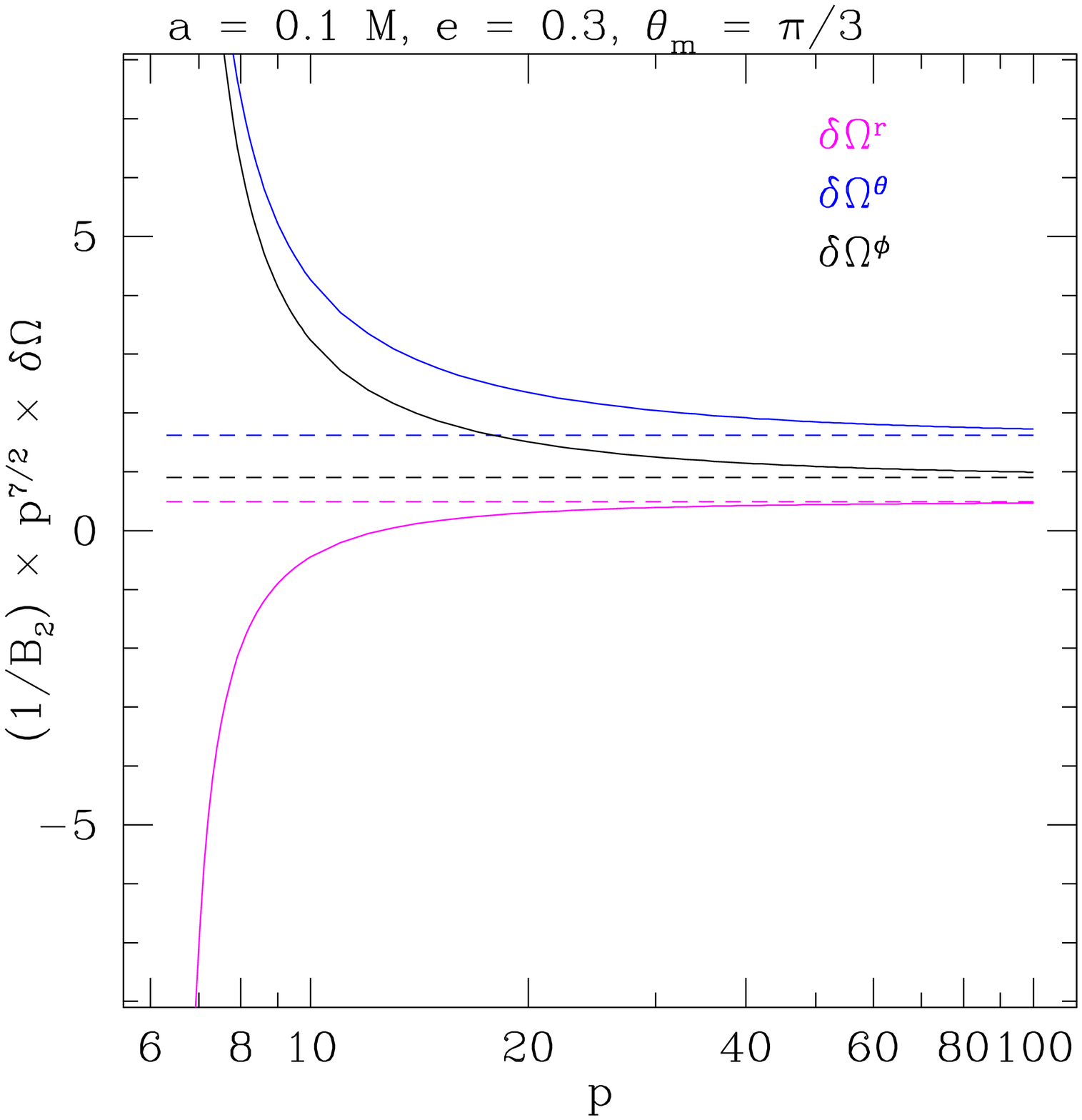}
\includegraphics[width=5.8cm]{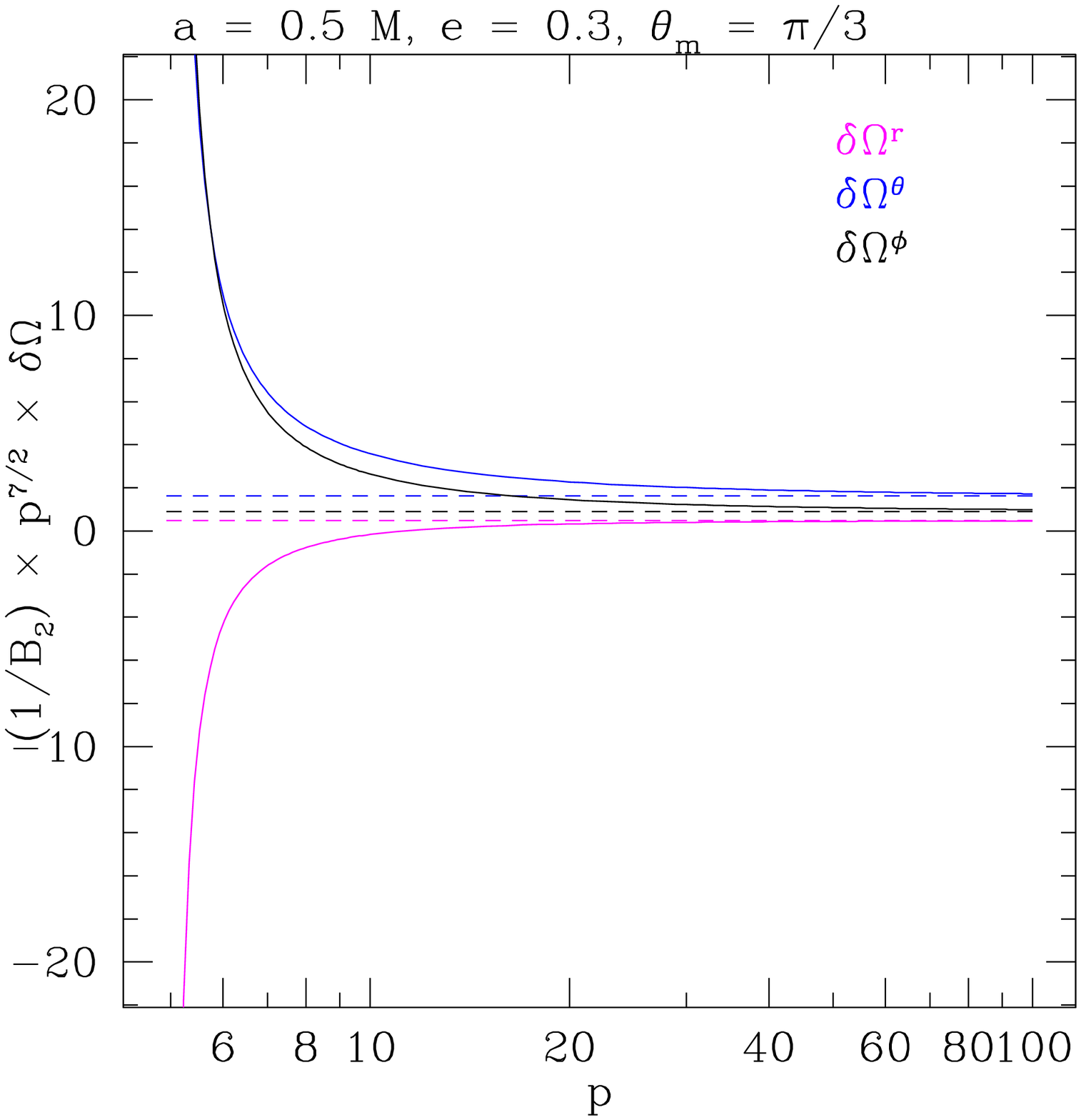}
\includegraphics[width=5.8cm]{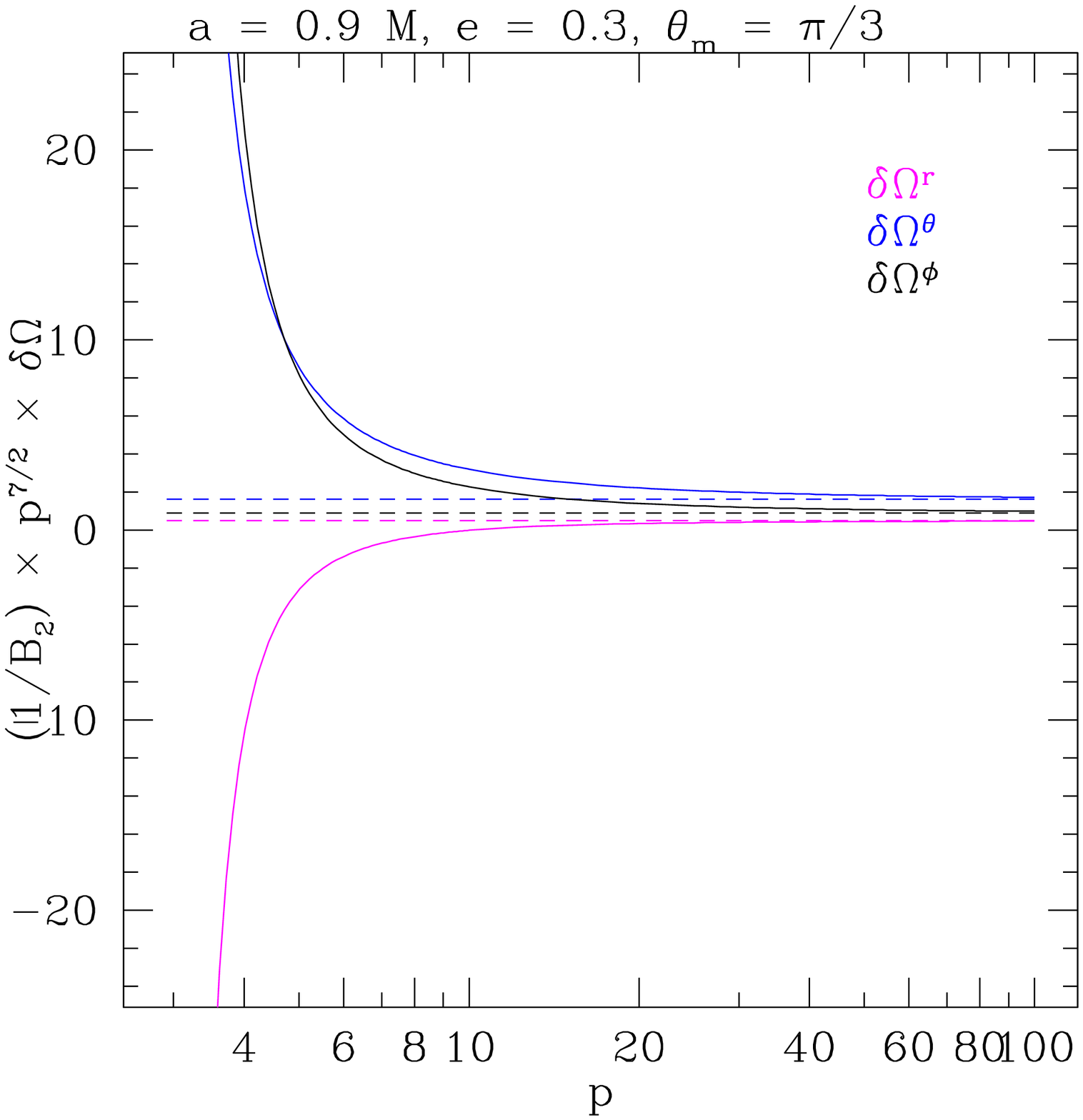}
\caption{Shifts to Kerr black hole orbital frequencies for an $l = 2$
bump.  As with the Schwarzschild results presented in Fig.\
{\ref{fig:dOm_l2}}, the shifts $\delta\Omega^{r,\theta,\phi}$ are
normalized by the bumpiness $B_2$ and scaled by $p^{7/2}$.  Rather
than examining a variety of orbital geometries, we here examine a few
black hole spins, showing results for $a = 0.1M$, $a = 0.5M$, and $a =
0.9M$.  Qualitatively, the results are very similar to what we find
for the Schwarzschild case.  The major difference is that the last
stable orbit is located at smaller $p$, so that these orbits can get
deeper into the strong field.  The overall impact of the bumps is
greater in these cases which reach deeper into the strong field.}
\label{fig:dOm_l2_Kerr}
\end{figure}

Figure {\ref{fig:dOm_l2_Kerr}} shows how an $l = 2$ bump changes Kerr
orbital frequencies.  We focus here on how black hole spin affects our
results, presenting results for a single orbit geometry ($e = 0.3$,
$\theta_m = \pi/3$).  The major impact of black hole spin is to change
the value of $p$ at which orbits become unstable.  For large spin,
orbits can reach deeper into the strong field, accumulating larger
anomalous shifts to their orbital frequencies.  Aside from this
behavior, spin has relatively little effect on the shifts: the three
panels are similar to one another and to the Schwarzschild result
(compare left-most panel of Fig.\ {\ref{fig:dOm_l2}}).  Similar
results hold for other orbital geometries, so we confine our plots to
these results.

\subsection{Octupole bumps ($l=3$)}

For the $l=3$ Kerr bump, we begin with Eq.\ (\ref{l3_potential}), then
follow the same procedure to take it to Boyer-Linquist coordinates as
described for the $l = 2$ Kerr bump.  The result is
\begin{eqnarray}
\psi_1^{l=3}(r, \theta) &=& \frac{B_3M^4}{4}\sqrt{\frac{7}{\pi}}
\frac{1}{d(r,\theta,a)^4}
\left[\frac{5L(r,\theta,a)^3\cos^3\theta}{d(r, \theta, a)^3} -
\frac{3L(r,\theta,a)\cos\theta}{d(r, \theta, a)} \right] \;,
\\
\gamma_1^{l=3}(r, \theta) &=& \frac{B_3M^5}{2}\sqrt{\frac{7}{\pi}}
\cos\theta \left[\frac{c_{30}(r,a) + c_{32}(r,a)\cos^2\theta +
c_{34}(r,a)\cos^4\theta + c_{36}(r,a)\cos^6\theta}{d(r,\theta,a)^7}
\right]\;,
\end{eqnarray}
where
\begin{eqnarray}
c_{30}(r,a) &=& -3r(r - 2M)\;,
\label{eq:g30_kerr}\\
c_{32}(r,a) &=& 10r(r - 2M) + 2M^5 - 3a^2\;,
\label{eq:g32_kerr}\\
c_{34}(r,a) &=& -7r(r-2M) + 10a^2\;,
\label{eq:g34_kerr}\\
c_{36}(r,a) &=& -2M^2 - 7a^2\;.
\label{eq:g36_kerr}
\end{eqnarray}

As with the Schwarschild $l = 3$ bumps, $\psi_1^{l=3}$ and
$\gamma_1^{l=3}$ are proportional to $\cos\theta$, so that
$\langle\Hcal_1\rangle = 0$.  Thus, for Kerr as for Schwarzschild,
there is no secular shift to orbital frequencies for $l = 3$, or any
other odd value of $l$.  We emphasize again that there will be
non-secular shifts to the motion which our orbit-averaged approach
misses by construction, and that it would be worthwhile to investigate
their importance in future work.

\subsection{Hexadecapole bumps ($l=4$)}

For $l = 4$, we begin with Eq.\ (\ref{l4_potential}).  Repeating our
procedure to take this into a Kerr bump, we find
\begin{eqnarray}
\psi_1^{l=4}(r,\theta) &=& \frac{B_4M^5}{16}\sqrt{\frac{9}{\pi}}
\frac{1}{d(r,\theta,a)^5}\left[\frac{35L(r,\theta,a)^4\cos^4\theta}
{d(r,\theta,a)^4} -
\frac{30L(r,\theta,a)^2\cos^2\theta}{d(r,\theta,a)^2} + 3\right]\;,
\\
\gamma_1^{l=4}(r, \theta) &=& B_4\sqrt{\frac{9}{\pi}} \left[
\frac{L(r,\theta,a)}{8} \frac{\left[c_{40}(r,a) +
c_{42}(r,a)\cos^2\theta + c_{44}(r,a)\cos^4\theta +
c_{46}(r,a)\cos^6\theta +
c_{48}(r,a)\cos^8\theta\right]}{d(r,\theta,a)^9} - 1\right]\,,
\nonumber\\
\end{eqnarray}
where
\begin{eqnarray}
c_{40}(r,a) &=& 8(r-M)^8 - 36M^2(r-M)^6 + 63M^4(r-M)^4 - 50M^6(r-M)^2 +
15M^8\;,
\label{eq:g40_kerr}\\
\nonumber c_{42}(r,a) &=& 36M^2(r-M)^6 - 126M^4(r-M)^4 + 120M^6(r-M)^2 - 30M^8 \\
		&& + 2a^2 \left[ 16(r-M)^6 - 54M^2(r-M)^4 + 63M^4(r-M)^2 - 25M^6 \right] \;,
\label{eq:g42_kerr}\\
c_{44}(r,a) &=& 63M^4(r-M)^4 - 70M^6(r-M)^2 + 15M^8
\nonumber\\
& & +
3a^2\left[4M^2\left(9(r-M)^4 - 21M^2(r-M)^2 + 10M^4\right) +
a^2\left(16(r-M)^4 - 36M^2(r-M)^2 + 21M^4\right)\right]\;,
\nonumber\\
\label{eq:g44_kerr}\\
c_{46}(r,a) &=& 2a^2\left[2a^4\left(8(r-M)^2 - 9M^2\right) +
9a^2M^2\left(6(r-M)^2 - 7M^2\right) + 7M^4\left(9(r-M)^2 - 5M^2\right)
\right]\;,
\label{eq:g46_kerr}\\
c_{48}(r,a) &=& a^4\left(8a^4 + 36a^2M^2 + 63M^4\right)\;.
\label{eq:g48_kerr}
\end{eqnarray}
It is a straightforward exercise to numerically compute
$\delta\Omega^{r,\theta,\phi}$ using $\psi_1^{l=4}$ and
$\gamma_1^{l=4}$ and build the Kerr analogs to the results we show in
Fig.\ {\ref{fig:dOm_l4}}.  The results are not markedly different from
those for Schwarzschild, modulo the fact that the orbit can (especially for
large spin) reach deeper into the strong field and hence accumulate
more bump-induced anomalous precession.  Since there are no
particularly surprising features in these results compared to what we
have already shown, we will not show such plots.

\section{Summary and future work} 
\label{sec:summary}

This analysis significantly improves on the earlier presentation of
bumpy black holes given in CH04, producing bumps with well-behaved
strong-field structure, and extending the concept to Kerr black holes.
These extensions greatly expand the astrophysical relevance of these
spacetimes.  We have also demonstrated how Hamilton-Jacobi theory can
be applied to orbits in bumpy spacetimes to categorize the anomalous
precessions arising from their bumps in a reasonably straightforward
manner.  Bumpy black holes can now, at least in principle, be used as
the foundation for strong-gravity tests with astrophysical data.

It's worth re-emphasizing why we propose to use bumpy black holes,
rather than using exact solutions which include black holes as a limit
(for example, the Novikov-Manko spacetime {\cite{mn92}} used in Ref.\
{\cite{glm08}}).  In large part, our choice is a matter of taste.  Our
goal is to tweak a black hole's moments in an arbitrary manner, so
that the non-Kerr nature is entirely under our control.  From the
standpoint of formulating a null experiment, it arguably makes no
difference whether the Kerr deviation takes one particular
form or another.  {\it Any} falsifiable non-Kerr form is good enough
to formulate the test.  To our minds, the nice feature of this
approach is that if, for example, a theory of gravity specifies that a
black hole should have the same moment structure as general relativity
up to some $l = L$, it is simple to design a spacetime tailored to
testing that theory.  The case of Chern-Simons gravity discussed in
Ref.\ {\cite{csbh}} shows that this motivation is not merely academic,
but is motivated by plausible alternatives to general relativity.

This analysis provides a complete description of ``mass-type'' bumps,
i.e., perturbations to the mass moments $M_l$.  We have not examined
``current-type bumps,'' shifts to the spin moments $S_l$.  It is also
worth bearing in mind that a pure multipole in the Weyl sector,
$\psi^l_1 \propto Y_{l0}(\theta)$ does not correspond to a shift in a
pure Geroch-Hansen multipole.  A companion paper by one of us (SJV)
addresses these issues {\cite{vigeland}}.  That analysis shows how
spin moments can be adjusted from their Kerr values by perturbing the
Ernst potential describing the spacetime {\cite{ernst}}.  It also
shows that a pure Weyl-sector multipole $\psi^l_1$ changes no
Geroch-Hansen moments {\it lower} than $l$.  As such, a pure
Geroch-Hansen moment perturbation can be assembled by combining
multiple Weyl-sector multipole perturbations.

Aside from these more formal issues, the next major step in this
program will be to use these foundations to formulate actual
strong-field gravity tests that can be applied to astrophysical data.
We imagine several directions that would be interesting to follow:

\begin{itemize}

\item {\it Extreme mass ratio inspiral (EMRI)}: The capture and
inspiral of stellar mass compact into massive black holes at galaxy
centers is one of the original motivations of this work.  Much of the
recent literature on testing and mapping black hole spacetimes has
centered on understanding the character of orbits in non-Kerr black
hole candidate spacetimes {\cite{brink1,brink2,haris}}, with an eye on
application to gravitational-wave measurement of EMRIs.

The full analysis of EMRIs in non-Kerr spacetimes is, in principle,
quite complicated since their non-Kerr-ness breaks the Petrov Type-D
character of Kerr black holes.  As such, it may be quite difficult to
accurately compute their radiation emission.  It may not be quite so
difficult in bumpy black hole spacetimes.  Thanks to the smallness of
their non-Kerr character, it may be fruitful to use a ``hybrid''
approach in which the short timescale motion is computed in the bumpy
spacetime, but the radiation generation and backreaction is computed
in the Kerr spacetime; a similar idea was suggested in the context of
modeling EMRI events in Chern-Simons gravity {\cite{ys09}}.  Given
that our goal is to formulate a null experiment, this hybrid may be
good enough for a useful test, in lieu of solving the entire radiation
reaction problem in non-Kerr spacetimes.

\item {\it Black hole-pulsar systems}: One of the goals of the planned
Square Kilometer Array (SKA) {\cite{smits}} is the discovery of a
black hole-pulsar binary system.  If such a system is discovered,
detailed observation over many years should be able to tease the
multipole structure of the black hole from the data.  Similar observations
of neutron star-pulsar binary systems have already allowed us to make
exquisite measurements of neutron star properties and
gravitational-wave emission {\cite{stairs_lr}}.  The tools developed
here may already be adequate for doing this analysis since such
binaries will have a relatively slow inspiral time.

\item {\it Accretion flows on black hole candidate}: Programs to
observe the (presumed) black hole at the center of our galaxy are
maturing very quickly; programs to study accretion flows onto stellar
mass black holes in x-rays are already quite mature.  In our galactic
center, the most precise measurements come from millimeter wavelength
radio emission from gas accreting onto this central object.  The
precision of these measurements is increasing to the point where we
will soon be able to use them to map the detailed strong-field
spacetime structure of the spacetime near Sagittarius A*.  It would be
a worthwhile exercise to repeat analyses of the appearance of these
flows in Kerr spacetimes {\cite{avery}} for bumpy spacetimes to see
how accurately such measurements may be able to probe the central
object's multipoles.  Such an analysis will require developing imaging
maps of bumpy black holes, going beyond the orbit frequency analysis
we have given here.  In the domain of x-rays, it would be very
interesting to extend work on quiescent accretion flows (as in, for
example, Ref.\ {\cite{nms08}}) to include accretion in bumpy black
hole spacetimes.  It may be possible to extend those analyses to fit
for multipoles beyond the black hole's mass and spin.

\end{itemize}

\acknowledgments

We thank Dimitrios Psaltis for useful feedback on bumpy black holes in
the early stage of this project, and for very helpful discussions
about testing black hole spacetimes with observations at the galactic
center.  We also thank Avi Loeb for discussion about galactic center
observations, Ilya Mandel for helpful advice and pointers to the
literature as we were completing this paper, and Tim Johannsen and
Nico Yunes for useful feedback on a preprint of this paper.  The
package {\sc Mathematica} was used to facilitate many of our
calculations, both analytic and numerical.  This work was supported by
NSF Grant PHY--0449884, and by NASA Grants NNG05G105G and NNX08AL42G;
SAH in addition gratefully acknowledges the support of the Adam J.\
Burgasser Chair in Astrophysics at MIT in completing this analysis.

\appendix

\section{Averaging functions along black hole orbits}
\label{app:averaging}

Key to computing the shifts that a black hole's bumps impart to its
orbital frequencies is averaging the Hamiltonian perturbation $\Hcal_1
= \Hcal_1(r,\theta)$ along the orbit.  The averaging we wish to use is
a time average:
\begin{equation}
\langle\Hcal_1\rangle = \lim_{T\to\infty}\frac{1}{2T}\int_{-T}^T
\Hcal_1\left[r(t),\theta(t)\right]\,dt\;.
\end{equation}
Implementing this integral is somewhat involved, so we review the
procedure here.  These details were first developed by Drasco and
Hughes (Ref.\ {\cite{dh04}}), hereafter DH04.

Most GR textbooks (e.g., Ref.\ {\cite{mtw}}, Chap.\ 33) describe the
motion of a small test body orbiting a Kerr black hole using Eqs.\
(\ref{eq:r_of_tau}) -- (\ref{eq:t_of_tau}).  Although the coordinate
motions are formally separated in these equations, in a practical
application they are not quite ``separated enough'' thanks to the
factors of $\Sigma = r^2 + a^2\cos^2\theta$ which appear on the
left-hand side of these equations.  These factors couple the radial
and polar motion, and complicate averaging functions of the form
$f(r,\theta)$ that are computed on an orbit.  Unless the $r$ and
$\theta$ periods coincide (or are in an integer ratio) one cannot
easily average over the $r$ and $\theta$ motion.

This residual coupling is eliminated by parameterizing the orbits
using what is now often called ``Mino time'' $\lambda$, defined by
$d\lambda = d\tau/\Sigma$.  Equations (\ref{eq:r_of_tau}) --
(\ref{eq:t_of_tau}) become
\begin{eqnarray}
\left(\frac{dr}{d\lambda}\right)^2 &=& R(r)\;,
\label{eq:r_of_lambda}\\
\left(\frac{d\theta}{d\lambda}\right)^2 &=& \Theta(\theta)\;,
\label{eq:theta_of_lambda}\\
\frac{d\phi}{d\lambda} &=& \Phi(r,\theta)\;,
\label{eq:phi_of_lambda}\\
\frac{dt}{d\lambda} &=& T(r,\theta)\;.
\label{eq:t_of_lambda}
\end{eqnarray}
Since this time variable explicitly separates the $r$ and $\theta$
motion, it is simple to construct $r(\lambda)$ using Eq.\
(\ref{eq:r_of_lambda}), and likewise to construct $\theta(\lambda)$
with (\ref{eq:theta_of_lambda}).  It is also simple to compute the $r$
and $\theta$ periods in Mino time: By inspection of Eqs.\
(\ref{eq:r_of_lambda}) and (\ref{eq:theta_of_lambda}), and taking into
account symmetries of the motion and the appropriate functions, we
have
\begin{eqnarray}
\Lambda^r &=& 2\int_{r_p}^{r_a} \frac{dr}{\sqrt{R(r)}}\;,
\label{eq:Lambda_r}\\
\Lambda^\theta &=& 4\int_{\theta_{\rm min}}^{\pi/2}
\frac{d\theta}{\sqrt{\Theta(\theta)}}\;.
\label{eq:Lambda_theta}
\end{eqnarray}
Following DH04, we define frequencies conjugate to these periods,
\begin{equation}
\Upsilon^{r,\theta} = 2\pi/\Lambda^{r,\theta}\;,
\end{equation}
and then introduce angles
\begin{equation}
w^{r,\theta} = \Upsilon^{r,\theta} \lambda\;.
\end{equation}
We then write the radial motion as a function of $w^r$,
\begin{equation}
r(w^r) = r(\lambda=w^r/\Upsilon^r)\;;
\end{equation}
we likewise parameterize the polar motion using $w^\theta$.  The key
concept behind the averaging is that we now allow the angles $w^r$ and
$w^\theta$ to {\it separately} vary.  This allows us to separately
average the $r$ and $\theta$ motions.

We are nearly ready to use these tools to average our Hamiltonian.
Before doing so, we define $\Upsilon^t$, the time-function
$T(r,\theta)$ averaged over the angles $w^r$ and $w^\theta$:
\begin{equation}
\Upsilon^t \equiv \frac{1}{(2\pi)^2}\int_0^{2\pi} dw^r\int_0^{2\pi}
dw^\theta\, T\left[r(w^r), \theta(w^\theta)\right]\;.
\label{eq:Upsilon_t}
\end{equation}
[In DH04, this quantity was denoted by a capital gamma; we are
adjusting the notation slightly to avoid conflict with $\Gamma$ as
defined in Eq.\ (\ref{Kerr_exact_Gamma}).  It is something of an abuse
of our notation, since the time $t$ is not periodic and $\Upsilon^i$
typically denotes a frequency.]

As shown in DH04, it is now simple to compute the long-time average of
any black hole orbit functional:
\begin{eqnarray}
\langle f \rangle &\equiv& \lim_{T\to\infty}\frac{1}{2T}\int_{-T}^T
f\left[r(t),\theta(t)\right]\,dt \nonumber\\ &=&
\frac{1}{\Upsilon^t(2\pi)^2}\int_0^{2\pi} dw^r\int_0^{2\pi} dw^\theta
f\left[r(w^r), \theta(w^\theta)\right] T\left[r(w^r),
\theta(w^\theta)\right]\;.
\label{eq:Kerraveraging}
\end{eqnarray}
This is the procedure we use to compute $\langle\Hcal_1\rangle$ for
all the computations presented in this paper.

\section{Newtonian precession frequencies}
\label{app:newton}

For weak-field orbits, we expect that the bumpy black hole frequency
shifts [Eqs.\ (\ref{eq:gen_shifts}) -- (\ref{eq:freq_shift})] are well
described using Newtonian gravity.  In this appendix, we compute the
relevant Newtonian frequency shifts; in Secs.\
{\ref{sec:results_schw}} and {\ref{sec:results_kerr}}, we show that
our general formulas limit to the results we develop here for large
radius orbits.

As in our relativistic calculation, we compute frequency shifts due to
multipolar ``bumps'' by examining the variation of a perturbed
Hamiltonian with respect to an orbit's action variables:
\begin{equation}
m\,\delta\Omega^i = \frac{\partial\langle\Hcal_1\rangle}{\partial J_i}\;.
\end{equation}
(Since there is no distinction between coordinate and proper time in
Newtonian gravity, these $\delta\Omega^i$ are the measurable
frequencies.)  The actions are defined just as in the relativistic
case, Eqs.\ (\ref{eq:Jr}) -- (\ref{eq:Jphi}).  For a body of mass $m$
orbiting a mass $M$ in Newtonian gravity, they become
\begin{eqnarray}
J_r &=& m\sqrt{pM}\left(\sqrt{\frac{1}{1 - e^2}} - 1\right)\;,
\\
J_\theta &=& m\sqrt{pM}\left(1 - \sin\theta_m\right)\;,
\\
J_\phi &=& m\sqrt{pM}\sin\theta_m\;.
\end{eqnarray}
The perturbation to the Hamiltonian for orbits is proportional to the
perturbation to the potential:
\begin{equation}
\Hcal_1^{\rm Newt} = m\delta V_l(r,\theta) = \frac{m B_l
M^{l+1}}{r^{l+1}}Y_{l0}(\cos\theta)\;.
\end{equation}
To perform the averaging, we first reparameterize both the radial and
angular motion in a manner similar to what we use for black hole
orbits:
\begin{eqnarray}
r &=& \frac{pM}{1 + e\cos\psi_r}\;,
\\
\cos\theta &=& \cos\theta_m\cos\left(\psi_r - \chi_0\right)\;.
\end{eqnarray}
Notice that the radial and angular motions vary in phase with one
another in the Newtonian limit: both complete a full cycle as the
angle $\psi_r$ varies from $0$ to $2\pi$.  The angle $\chi_0$ is an
offset phase between these motions.  Inserting this parameterization
into the Newtonian equations of orbital motion, we find
\begin{equation}
\frac{d\psi_r}{dt} = \sqrt{\frac{M}{p^3}}\left(1 +
e\cos\psi_r\right)^2\;.
\end{equation}
The averaged Hamiltonian is then given by
\begin{eqnarray}
\langle\Hcal_1^{\rm Newt}\rangle &=&
\frac{m}{T_K}\int_0^{T_K} \delta V_l\left[r(t),\theta(t)\right]dt
\nonumber\\
&=&\frac{m}{T_K}\int_0^{2\pi}
\left(\frac{d\psi_r}{dt}\right)^{-1} \delta
V_l\left[r(\psi_r),\cos\theta(\psi_r)\right]d\psi_r\;,
\end{eqnarray}
where $T_K = 2\pi M\sqrt{p^3/(1 - e^2)^3}$ is the Keplerian orbital
period.  [To put this in a more familiar form, recall that the orbit's
semi-major axis $A = pM/(1 - e^2)$.]

Using these results, we now examine the Newtonian limit of $l = 2$,
$3$, and $4$ black hole bumps.

\subsection{Quadrupole ($l = 2$)}

The quadrupole bump is given by the potential
\begin{equation}
\delta V^{l=2} = \frac{B_2M^3}{4r^3}\sqrt{\frac{5}{\pi}}
\left[3\cos^2\theta - 1\right]\;,
\label{eq:V2_Newt}
\end{equation}
for which we find
\begin{equation}
\langle\Hcal_1\rangle = \frac{mB_2}{8p^3}\sqrt{\frac{5}{\pi}} \left(1
- e^2\right)^{3/2} \left(1 - 3\sin^2\theta_m\right)\;.
\end{equation}
Varying this averaged Hamiltonian with respect to $J_{r,\theta,\phi}$,
we find
\begin{eqnarray}
\delta\Omega^r &=&
\frac{3B_2}{8M}\frac{1}{p^{7/2}}\sqrt{\frac{5}{\pi}}(1 - e^2)^2
\left(3\sin^2\theta_m-1\right)\;,
\label{eq:deltaOmega_r_l2_Newt}\\
\delta\Omega^\theta &=&
\frac{3B_2}{8M}\frac{1}{p^{7/2}}\sqrt{\frac{5}{\pi}}(1 - e^2)^{3/2}
\left[\sin^2\theta_m\left(5 + 3\sqrt{1 - e^2}\right) - \sqrt{1 - e^2}
- 1\right]\;,
\label{eq:deltaOmega_th_l2_Newt}\\
\delta\Omega^\phi &=&
\frac{3B_2}{8M}\frac{1}{p^{7/2}}\sqrt{\frac{5}{\pi}}(1 - e^2)^{3/2}
\left[\sin^2\theta_m\left(5 + 3\sqrt{1 - e^2}\right) - 2\sin\theta_m -
\sqrt{1 - e^2} - 1\right]\;.
\label{eq:deltaOmega_ph_l2_Newt}
\end{eqnarray}
These frequencies are written using the Kepler frequency $\omega_K =
2\pi/T_K$.

Equations (\ref{eq:deltaOmega_r_l2_Newt}) --
(\ref{eq:deltaOmega_ph_l2_Newt}) reproduce well-known results for
motion in a spherical potential augmented by a quadrupole
perturbation.  To facilitate comparison with the literature, it is
useful to change our description of the orientation of the orbital
plane from $\theta_m$ (the minimum angle $\theta$ reached over an
orbit) to $\iota = \pi/2 - \theta_m$ (the orbit's inclination with
respect to the equatorial plane).  We then construct the precession
frequencies
\begin{eqnarray}
\Omega^{\rm apsis} &=& \delta\Omega^\theta - \delta\Omega^r =
\omega^K\times\frac{3B_2}{8M}\frac{1}{p^2}\sqrt{\frac{5}{\pi}}
\left(5\cos^2\iota - 1\right)\;,
\label{eq:Newtonianl2_apsidal}
\\
\Omega^{\rm plane} &=& \delta\Omega^\phi - \delta\Omega^\theta =
-\omega^K\times\frac{3B_2}{4M}\frac{1}{p^2}\sqrt{\frac{5}{\pi}}
\cos\iota\;;
\label{eq:Newtonianl2_plane}
\end{eqnarray}
$\Omega^{\rm apsis}$ describes the frequency of the orbit's apsidal
precession within its orbital plane, and $\Omega^{\rm plane}$ the
frequency at which the orbital plane precesses around the symmetry
axis.  These frequencies reproduce expressions that can be found in
the literature; cf.\ Sec.\ 12.3C of Ref.\ {\cite{goldstein}}, and
Sec.\ 3.6.2 of Ref.\ {\cite{capderou}}.

The apsidal precession (\ref{eq:Newtonianl2_apsidal}) is due to
beating between the orbit's radial and polar motions.  In many
relativity applications, one is interested in the precession of an
orbit's periastron, which is due to a beat between the orbit's radial
and azimuthal motions.  For the quadrupole bump, we find
\begin{equation}
\Omega^{\rm peri} = \delta\Omega^\phi - \delta\Omega^r = \omega_K
\times \frac{3B_2}{8M}\frac{1}{p^2} \sqrt{\frac{5}{\pi}}
\left(5\cos^2\iota - 2\cos\iota - 1\right)\;.
\end{equation}
Taking the equatorial limit ($\iota = 0$), we see that this result
agrees with Eq.\ (A4) of CH04 provided we identify their parameter $Q$
with $B_2M^3\sqrt{5/4\pi}$.  Comparison of the potential used in CH04
[their Eq.\ (A2)] with our form [Eq.\ (\ref{eq:V2_Newt})] shows that
this identification is exactly correct.

\subsection{Octupole shift ($l = 3$)}

The octupole bump is given by the potential
\begin{equation}
\delta V^{l=3} = \frac{B_3M^4}{4r^4}\sqrt{\frac{7}{\pi}}
\left[5\cos^3\theta - 3\cos\theta\right]\;,
\end{equation}
leading to
\begin{equation}
\langle\Hcal_1\rangle = \frac{3mB_3e}{16p^4}\sqrt{\frac{7}{\pi}}
\left(1 - e^2\right)^{3/2}\cos\theta_m \left(5\cos^2\theta_m - 4\right)
\cos\chi_0\;.
\end{equation}
This is proportional to $\cos\chi_0$, and so depends on the phase
offset of the radial and angular motions.  Over very long timescales,
precessions will average out this dependence.  However, on timescales
that are not long enough for $\chi_0$ to vary over its full range,
there is a residual impact, leading to octupolar shifts to observable
frequencies.  This can be very important in Newtonian celestial
mechanics.  For black hole applications, where the $r$ and $\theta$
motions do not vary in phase with one another, this averaging will be
much stronger, and treating $\langle{\cal H}_1\rangle \simeq 0$ should
be much more accurate.

Varying the averaged octupole Hamiltonian shift, we find
\begin{eqnarray}
\delta\Omega^r &=&
\frac{3B_3}{16eM}\frac{1}{p^{9/2}}\sqrt{\frac{7}{\pi}}(1 - e^2)^2(1 - 4e^2)
\left(5\cos^2\theta_m - 4\right)\cos\chi_0\;,
\label{eq:deltaOmega_r_l3_Newt}\\
\delta\Omega^\theta &=&
\frac{3B_3}{16eM}\frac{1}{p^{9/2}}\sqrt{\frac{7}{\pi}}(1 - e^2)^{3/2}
\left[\cos\theta_m\left[4\left(1 - \sqrt{1 - e^2} + 4e^2\left(1 + \sqrt{1 -
e^2}\right)\right) + 15e^2\sin^2\theta_m\right]\right.
\nonumber\\
& &\left.-4e^2\sin\theta_m\tan\theta_m -5\left[1 - \sqrt{1 - e^2} +
4e^2\left(1 + \sqrt{1 - e^2}\right)\right]\cos^3\theta_m\right]\cos\chi_0\;,
\label{eq:deltaOmega_th_l3_Newt}\\
\delta\Omega^\theta &=&
\frac{3B_3}{16eM}\frac{1}{p^{9/2}}\sqrt{\frac{7}{\pi}}(1 - e^2)^{3/2}
\left[\cos\theta_m\left[4 - 4\sqrt{1 - e^2} + 16e^2\sqrt{1 - e^2} +
e^2\left[16 + 15\left(\sin\theta_m - 1\right)\right]\sin\theta_m\right]
\right.
\nonumber\\
& &\left.- 4e^2(\sin\theta_m - 1)\tan\theta_m - 5\left[1 - \sqrt{1 -
    e^2} + 4e^2\left(1 + \sqrt{1 -
    e^2}\right)\right]\cos^3\theta_m\right]\cos\chi_0\;.
\label{eq:deltaOmega_ph_l3_Newt}
\end{eqnarray}
As expected from the form of the averaged Hamiltonian, all precession
frequencies are proportional to the cosine of the offset phase
$\chi_0$.  {\it On average} we therefore find no influence from the $l
= 3$ perturbation, nor from any odd $l$ multipolar bump.  The
``instantaneous'' impact of odd $l$ bumps is, however, non-zero.

\subsection{Hexadecapole shift ($l = 4$)}

Finally, for the hexadecapole bump, we have
\begin{equation}
\delta V^{l=4} = \frac{B_4M^5}{16r^5}\sqrt{\frac{9}{\pi}}
\left[35\cos^4\theta - 30\cos^2\theta + 3\right]\;,
\end{equation}
leading to
\begin{equation}
\langle\Hcal_1\rangle = \frac{3mB_4}{256p^5}\sqrt{\frac{9}{\pi}}
\left(1 - e^2\right)^{3/2}\left[8(2 + 3e^2) - 20\cos^2\theta_m(4 +
6e^2 + 3e^2\cos2\chi_0) + 35\cos^4\theta_m(2 + 3e^2 +
2e^2\cos2\chi_0)\right]\;.
\end{equation}
We focus on the secular (long-time average) precessions, and average
this over $\chi_0$, leaving
\begin{equation}
\langle\Hcal_1\rangle = \frac{3mB_4}{256p^5}\sqrt{\frac{9}{\pi}}
\left(1 - e^2\right)^{3/2}(2 + 3e^2)\left(8 - 40\cos^2\theta_m +
35\cos^4\theta_m\right)\;.
\end{equation}
The precession frequencies which arise from this are
\begin{eqnarray}
\delta\Omega^r &=&
-\frac{45B_4}{256M}\frac{1}{p^{11/2}}\sqrt{\frac{9}{\pi}}(1 - e^2)^2
\left(8 - 40\cos^2\theta_m + 35\cos^4\theta_m\right)\;,
\label{eq:deltaOmega_r_l4_Newt}\\
\delta\Omega^\theta &=&
-\frac{15B_4}{256M}\frac{1}{p^{11/2}}\sqrt{\frac{9}{\pi}} (1 -
e^2)^{3/2} \left[8\left(8 + 3e^2\left(3 + \sqrt{1 - e^2}\right)\right)
  - 4\left(62 + e^2\left(63 + 30\sqrt{1 -
    e^2}\right)\right)\cos^2\theta_m\right.
\nonumber\\
& &\left. + 7\left(28 + 3e^2\left(9 + 5\sqrt{1 -
  e^2}\right)\right)\cos^4\theta_m\right]\;,
\label{eq:deltaOmega_th_l4_Newt}\\
\delta\Omega^\phi &=& \delta\Omega^\theta +
\frac{15B_4}{256M}\frac{1}{p^{11/2}}\sqrt{\frac{9}{\pi}} (1 -
e^2)^{3/2}(2 + 3e^2)\sin\theta_m\left(28 - 40\cos^2\theta_m\right)\;.
\label{eq:deltaOmega_ph_l4_Newt}
\end{eqnarray}

\end{document}